\documentclass{article}

\PassOptionsToPackage{numbers, compress}{natbib}

    \usepackage[final]{neurips_2024}

\usepackage[utf8]{inputenc} 
\usepackage[T1]{fontenc}    
\usepackage[colorlinks, citecolor=blue]{hyperref}       
\usepackage{url}            
\usepackage{booktabs}       
\usepackage{amsfonts}       
\usepackage{nicefrac}       
\usepackage{microtype}      
\usepackage{xcolor}         

\usepackage{amsmath}
\usepackage{amssymb}
\usepackage{amsthm}
\usepackage{graphicx}
\usepackage{bm}
\usepackage{bbm}
\usepackage{array} 

\newcommand{\zi}{z_{\text{i}}}
\newcommand{\zm}{z_{\text{m}}}
\newcommand{\tildezm}{\tilde{z}_{\text{m}}}
\newcommand{\zt}{z_{\text{t}}}

\newcommand{\Zm}{\mathcal{Z}_\text{m}}

\theoremstyle{definition}

\title{Diff4Steer: Steerable Diffusion Prior for Generative Music Retrieval with Semantic Guidance}

\author{%
  Xuchan Bao\thanks{Work done during internship at Google. Correspondence to judithyueli@google.com} \\
  University of Toronto\\
  \texttt{} \\
   \And
  Judith Yue Li\\
  Google Research\\
   \And
  Zhong Yi Wan\\
  Google Research\\
   \And
  Kun Su\\
  Google Research\\
   \And
  Timo Denk\\
  Google DeepMind\\
   \And
  Joonseok Lee\\
  Google Research\\
  Seoul National University\\
   \And
  Dima Kuzmin\\
  Google Research\\
  \And
  Fei Sha\\
  Google Research\\
}

\begin{document}

\maketitle

\vspace{-1em}
\begin{abstract}
Modern music retrieval systems often rely on fixed representations of user preferences, limiting their ability to capture users' diverse and uncertain retrieval needs. To address this limitation, we introduce Diff4Steer, a novel generative retrieval framework that employs lightweight diffusion models to synthesize diverse seed embeddings from user queries that represent potential directions for music exploration. Unlike deterministic methods that map user query to a single point in embedding space, Diff4Steer provides a statistical prior on the target modality (audio) for retrieval, effectively capturing the uncertainty and multi-faceted nature of user preferences. 
Furthermore, Diff4Steer can be steered by image or text inputs, enabling more flexible and controllable music discovery combined with nearest neighbor search. Our framework outperforms deterministic regression methods and LLM-based generative retrieval baseline in terms of retrieval and ranking metrics, demonstrating its effectiveness in capturing user preferences, leading to more diverse and relevant recommendations. Listening examples are available at tinyurl.com/diff4steer.
\end{abstract}
\vspace{-1em}

\section{Introduction}
Modern retrieval systems~\citep{Koren2009MatrixFT,Covington2016DeepNN}, including those for music~\citep{Kim2007AMR}, often employ embedding-based dense retrieval system for candidate generation. These systems use a joint embedding model~(JEM)~\citep{huang2022mulan,elizalde2022clap} to obtain deterministic representations of queries, known as seed embeddings, within a semantic space shared with the retrieval candidates. The seed embeddings provide the personalized starting point in the target embedding space for retrieving similar music via nearest neighbor search.

While JEM-based system provides computationally efficient retrieval solution, they are insufficient in modeling user's diverse and uncertain retrieval preference. First, JEM only supports users expressing music preference or steer the retrieval results via specific modalities that the JEM is built on. Moreover, music discovery is an inherently ambiguous task with many possible outcomes -- there is no one-to-one mapping between the query and seed embedding given the large uncertainty of how a user's music preference can be fully specified. For example, ``energetic rock music'' could mean ``punk rock'' for some, or ``hard rock'' for others. Modeling user preference using deterministic seed embedding can lead to monotonous and inflexible recommendations~\citep{anderson2020algorithmic}. In essence, for creative applications, it is crucial to explore users' possible intentions (by allowing them to steer the retrieval results via instructions) and to return relevant and diverse retrieval results.

\begin{figure}[t]
    \centering
    \includegraphics[width=1.0\linewidth]{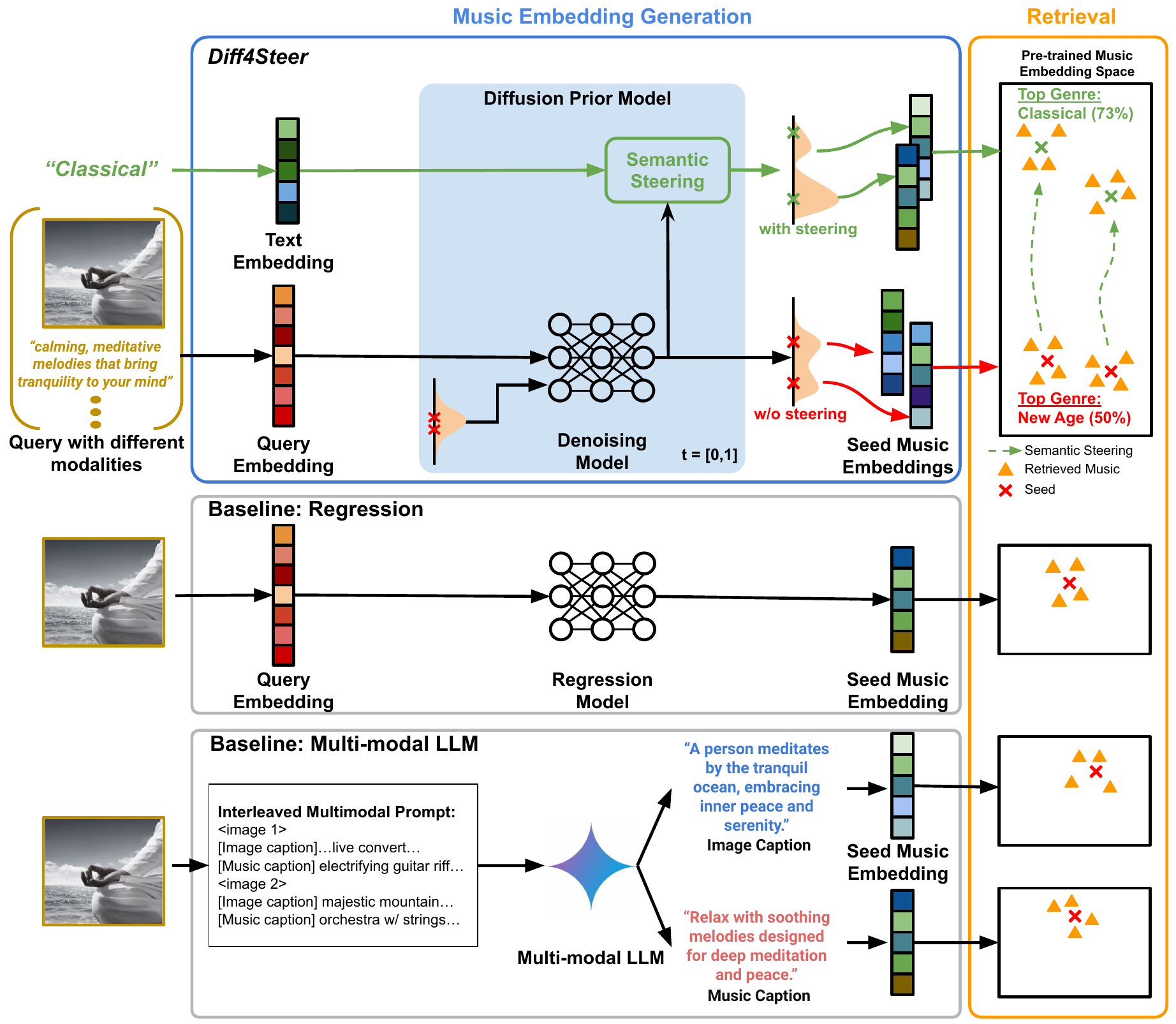}
    \caption{Overall diagram of our generative retrieval framework for cross-modal music retrieval, with comparison to the regression and multi-modal LLM baselines.
    }
    \label{fig:overall}
    \vspace{-1em}
\end{figure}

To better represent diversity and uncertainty in users' retrieval preference, we introduce a novel framework \textit{Diff4Steer} (Figure~\ref{fig:overall}) for music retrieval that leverages the strength of generative models for synthesizing potential directions to explore, The directions are implied by the generated seed embeddings: a collection of vectors in the music embedding space that represent the distribution of a user's music preferences given retrieval queries. Concretely, our lightweight diffusion-based generative models give rise to a statistical prior on the target modality -- in our case, audio -- for the music retrieval task. Furthermore, the prior can be conditioned on image or text inputs, to generate samples in the audio embedding space learned by the pre-trained joint embedding model. They are then used to retrieve the candidates using nearest neighbor search. Given that constructing a large-scale dataset that contains the aligned multimodal data (steering info, source modality, target modality) is very difficult, we also leverage the diffusion models' flexibility in sampling-time steering to incorporate additional text-based user preference specifications. This eliminates the need for expensive, data-hungry joint embedding training across all modalities. 

While we have seen that diffusion-based generative approaches~\citep{rombach2022high, schneider2023mo, liu2023audioldm} can ensure diversity and quality in the embedding generation, in this work we investigate their performance on retrieval tasks. We demonstrate that our generative music retrieval framework achieves competitive retrieval and ranking performance while introducing much-needed diversity. A comparison with deterministic regression methods shows that \textit{Diff4Steer} achieves superior retrieval metrics. This is thanks to the higher quality of the generated embedding, which reflects the underlying data distribution, as well as incorporating uncertainty in modeling user preferences.

\section{Approach}

\textbf{Music Embedding Diffusion Prior}
Following the EDM~\citep{karras2022elucidating} formulation, our diffusion prior is parametrized by a denoiser neural network $D(\tilde{z}_\text{m}, \sigma, q)$, which learns to predict the clean music embedding $z_\text{m}$ given a noisy embedding $\tilde{z}_\text{m}=\zm+\epsilon\sigma$, noise level $\sigma$ and cross-modal query $q$, by minimizing the $\ell_2$ loss:
\begin{equation}
\label{eq:denoising_loss}
    L(\theta; \mathcal{D}) = \mathbb{E}_{\epsilon\sim\mathcal{N}(0, 1), \sigma\sim \eta(\sigma), \{\zm, q\} \in \mathcal{D}} \left[\lambda(\sigma) \cdot \Vert D_\theta(\tilde{z}_\text{m}, \sigma, q) - \zm \Vert^2\right],
\end{equation}
where $\epsilon$ draws from a standard Gaussian, $\eta$ is a training distribution for $\sigma$, $\lambda$ is the loss weighting, and $\mathcal{D}$ denotes training dataset with paired $(\zm, q)$ examples. Sampling is performed by solving the stochastic differential equation (SDE)
\begin{equation}
\label{eq:sampling_sde}
    d\tilde{z}_{\text{m},t} = \left[\left(\dot{\sigma}_t/\sigma_t\right) \tilde{z}_{\text{m}, t} - 2(\dot{\sigma}_t/\sigma_t)D_\theta\left(\tilde{z}_{\text{m},t}, \sigma_t, q\right)\right]dt + \sqrt{2\dot{\sigma}_t\sigma_t}\;dW_t,
\end{equation}
from $t=1$ to $0$ with noise schedule $\sigma_t$ and initial condition $\tilde{z}_{\text{m},1}\sim\mathcal{N}(0, \sigma_{t=1})$, using a first-order Euler-Maruyama solver.

\textbf{Classifier-free Guidance (CFG)}  \citep{ho2022classifier} is used to enhance the alignment of the sampled music embeddings to the cross-modal inputs. 
During training, the condition $q$ is randomly masked with a zero vector with probability $p_{\text{mask}}$, such that the model simultaneously learns to generate conditional and unconditional samples with shared parameters.
At sampling time, the effective denoiser $D'_\theta$ is an affine combination of the conditional and unconditional versions
\begin{equation}
\label{eq:cfg}
    D'_\theta(\tildezm, \sigma, q) = (1 + \omega) D_\theta(\tildezm, \sigma, q) - \omega D_\theta(\tildezm, \sigma, \bm{0}),
\end{equation}
where $\omega$ denotes the CFG strength, which boosts alignment with $q$ when $\omega>0$. $\omega=-1.0$ indicates unconditional generation.

\textbf{Additional text steering}
can be applied when the underlying music embedding space is that of a text-music JEM~\citep{huang2022mulan}. In such case, the JEM provides a text encoder $E_\text{t}: T\rightarrow\zt$ with a text-music similarity measure via the vector dot product $\langle\zt,\zm\rangle$.
This allows us to incorporate (potentially multiple) text steering signals by modifying the denoising function \emph{at sampling time}:
\begin{equation}
\label{eq:jem_guidance}
    D''_\theta\left(\tildezm, \sigma, q, \left[z_{\text{t},1}, z_{\text{t},2}, ...\right]\right) = D'_\theta(\tildezm, \sigma, q) + \sum_n k_n \nabla_{\tildezm} \left\langle D'_\theta(\tildezm, \sigma, q), z_{\text{t},n}\right\rangle,
\end{equation}
where $k_n$ is the strength for the $n$-th text steer signal $z_{\text{t},n}$. Each $k_n$ can be positive/negative depending on whether features described by corresponding texts are desirable/undesirable in the samples. We note that such steering comes in addition to explicit condition $q$, which may itself contain text inputs.

\section{Experimental settings}\label{sec:exp-settings}






\subsection{Tasks and Datasets}
In our retrieval experiments, we use our diffusion prior model to perform several downstream tasks simultaneously, namely image-to-music retrieval, text-to-music retrieval and image-to-music retrieval with text steering. For image-to-music tasks the query embedding is CLIP~\citep{radford2021learning}. For text-to-music retrieval or text steering, text is encoded via MuLan text embedding and incorporated as a steering condition to steer the seed embedding generation using genre or music caption.

\textbf{YouTube 8M (YT8M)}~\citep{abu2016yt8m} is a dataset originally developed for the video classification task, equipped with video-level labels.
We use the 116K music videos in this dataset to generate (music, image) pairs by extracting 10s audios and randomly sampling a video frame in the same time window. This dataset is primarily used for training.

We use two other expert-annotated datasets for evaluation. First,
\textbf{MusicCaps (MC)}~\citep{agostinelli2023musiclm}
is a collection of 10s music audio clips with human-annotated textual descriptions. We extend the dataset with an image frame extracted from the corresponding music video.
\textbf{MelBench (MB)}~\citep{chowdhury2024melfusion}
is another collection of images paired with matching music caption and music audio annotated by music professionals.

\subsection{Model and training}
\label{sec:training-sampling}
We use a 6-layer ResNet with width of 4096 as the backbone of the denoising model.
For classifier-free guidance, we use a condition mask probability $p_\text{mask} = 0.1$, in order to simultaneously learn the conditional and unconditional denoising model under shared parameters. We train the denoising model on paired image and music embeddings from the YT8M music videos. We use the Adam~\citep{kingma2014adam} optimizer under cosine annealed learning rate schedule~\citep{loshchilov2016sgdr} with peak rate $10^{-5}$. Our model has 282.9M parameters in total and can fit into one TPU. We train our model for 2M steps, which takes around two days on a single TPU v5e device.

\subsection{Baselines}
\label{sec:exp-baselines}

\textbf{MuLan.}~\citep{huang2022mulan}
As a text-music JEM, MuLan enables text-to-music retrieval through a nearest neighbor search based on the dot product similarity between a text query and candidate music embeddings.

\textbf{Regression model.} 
We train a regression baseline model that maps the query embeddings (CLIP image embedding) to MuLan audio embeddings deterministically using the same architecture as the diffusion model (excluding noise).

\textbf{Multi-modal Gemini.} The multi-modal Gemini serves as a strong baseline for our image-to-music retrieval tasks. We leverage a few-shot interleaved multi-modal prompt that given an image it can generate image caption or matching music caption. Specifically, \textit{Gemini-ImageCap} encodes the generated image caption into a MuLan text embedding for retrieving candidate audio embeddings. \textit{Gemini-MusicCap} encodes the generated music caption into a MuLan text embedding for retrieving candidate audio embeddings.


\vspace{-0.5em}
\subsection{Evaluation Metrics}

\textbf{Embedding Quality.}
We use two metrics to measure the quality of generated music embeddings: Fr\'{e}chet MuLan Distance (FMD) and mean intra-sample cosine similarity (MISCS).
FMD is inspired by Fr\'{e}chet Inception Distance (FID)~\citep{heusel2017gans} and measures the similarity of a set of generated music embeddings to a population of real music embeddings in distribution.

\textbf{Music-image Alignment (M2I).} 
Assessing alignment between generated music embeddings and input images is challenging due to their distinct domains. Leveraging the shared text modality in CLIP and MuLan, we use text as a bridge for evaluating music-image (M2I) alignment following~\citet{chowdhury2024melfusion} . This approach eliminates the need for paired data and instead requires a set of images and a separate set of texts. By encoding texts into both CLIP and MuLan embeddings, M2I is calculated as the average of the product of two cosine similarities.

\textbf{Retrieval Metrics.}
We evaluate retrieval results using three metrics.
First, we report recall@K (R@K), a standard metric in information retrieval.
However, image-to-music or text-to-music retrieval is inherently subjective, 
often featuring one-to-many mappings.
Thus, recall@K alone is insufficient, and we also report diversity using mean intra-sample cosine similarity (MISCS) and triplet accuracy (TA) to provide a more comprehensive evaluation.



\vspace{-0.5em}
\section{Results and Discussion}\label{sec:results}
In this section, we present experimental results that demonstrate: (1) our \textit{Diff4Steer} model effectively generates high-quality seed embeddings; (2) \textit{Diff4Steer} achieves competitive retrieval performance compared to other cross-modal retrieval methods and significantly improves retrieval diversity, and enables effective and personalized steering of seed embeddings during inference.
\vspace{-0.5em}
\subsection{Quality of the Generated Seed Embeddings}
\label{sec:exp-embed}
Table~\ref{tab:fmd-diversity} presents a comparison of embedding quality between \textit{Diff4Steer} and the regression baseline for both image-to-music and text-to-music tasks across multiple datasets. Results show that our diffusion prior model consistently exhibits significantly lower FMD, indicating higher quality and greater realism in generated MuLan audio embeddings compared to the baseline. 
In addition, the diffusion model achieves significantly lower MISCS scores across all datasets, indicating that it allows us to generate diverse samples, which is impossible with a regression model. 

\begin{table}[t]
    \centering
    \caption{FMD and MISCS of the generated music embeddings for YT8M, MC and MB datasets (image2music). Across all the datasets, our diffusion model outperforms the deterministic model in both embedding quality (FMD) and diversity (MISCS).}
    \begin{tabular}{l|cccccc}\toprule
         Method & \multicolumn{3}{c}{FMD $\downarrow$} & \multicolumn{3}{c}{MISCS $\downarrow$}\\
         & YT8M & MC & MB & YT8M & MC & MB \\\midrule
        Regression & 0.480 & 0.507 & 0.518 & 1.000 & 1.000 & 1.000 \\ 
        \textbf{Diff4Steer (ours)} & \textbf{0.161} & \textbf{0.152} & \textbf{0.172} & \textbf{0.805} & \textbf{0.391} & \textbf{0.404} \\ 
        \bottomrule
    \end{tabular}
    \label{tab:fmd-diversity}
\end{table}

There is a dynamic relationship between classifier-free guidance (CFG) strength $\omega$ and the quality and diversity of embeddings generated by our diffusion model.
With a guidance strength of $\omega=-1.0$, corresponding to unconditional samples, FMD initially deteriorates, then improves, and eventually gets worse again with excessively high $\omega$.
Conversely, diversity consistently decreases with increasing $\omega$, highlighting the inherent trade-off between embedding quality and diversity.



\subsection{Embedding-based Music Retrieval}
We show embedding-based music retrieval results in Table~\ref{tab:img-retrieval}.
The image CFG strength is an important hyperparameter, and we tune it using the FMD score, based on the YT8M evaluation split. For the remaining evaluations in this paper, we set the image guidance strength to be 19.0.

\label{sec:exp-retrieval}



\textbf{High-quality embeddings leads to high recall.} 
A key finding from Table~\ref{tab:img-retrieval} is that our \textit{Diff4Steer} model has significantly higher recall and triplet accuracy, compared to the regression and multi-modal Gemini baselines. This underscores the value of our approach for music retrieval applications. Notably, while the regression model has the highest M2I in the image-to-music task, it falls short in standard retrieval metrics. This observation, along with the FMD results in Section~\ref{sec:exp-embed}, highlights the crucial role of high-quality seed embeddings in achieving optimal retrieval performance.

\textbf{Modality gap may harm retrieval results.} 
For the multi-modal Gemini baselines, the image-to-music embedding generation is broken down to multiple stages. We use text (image or music captions) as an intermediate modality, thereby introducing potential modality gap. As shown in Table~\ref{tab:img-retrieval}, despite the power of the general-purpose LLMs, multi-modal Gemini baselines have worse retrieval performance than our \textit{Diff4Steer} model, likely due to the loss of information with the modality gap. Additionally, our model offers a significantly lighter weight solution in terms of training consumption and latency compared to multi-modal foundation models. 

\textbf{One model for all modality.}
Notably, our \textit{Diff4Steer} model demonstrates competitive performance on genre-to-music and caption-to-music retrieval tasks (the second and third groups in Table~\ref{tab:img-retrieval}) despite not being trained on paired text and music data. This is achieved by unconditionally generating audio embeddings guided by text-music similarity. Compared to the regression baseline, \textit{Diff4Steer} achieves superior results on most retrieval and ranking metrics, especially on the tasks that involve higher retrieval uncertainty, \textit{e.g.}, genre-to-music retrieval.



\textbf{Text steering improves recall.}
Furthermore, we explore the extent to which text steering helps with retrieval. In addition to the image input, we provide our diffusion model with the genre label or ground truth caption at inference time. The last group in Table~\ref{tab:img-retrieval} shows that when steered with the additional textual information, the models achieve significantly higher recall and triplet accuracy.




\begin{table}[t]
    \centering
    \caption{Music retrieval results of our model and various baselines, evaluated on MC and MB.}
    \label{tab:img-retrieval}
    {\setlength\tabcolsep{2.7pt}
    \begin{tabular}{lccccccccc}
        \toprule
        Method & Input& \multicolumn{4}{c}{MC w/ Images} & \multicolumn{4}{c}{MB} \\
        &  & R@100 & R@10 & M2I & TA &R@100 & R@10 & M2I & TA \\
        \midrule
        Gemini-ImageCap & image & 0.215 & 0.055 & 89.12& 0.488 & 0.162 &  0.036 & 90.32& 0.685\\
        Gemini-MusicCap & image & 0.210 & 0.049 &84.48 & 0.521 & 0.145 &  0.026 &88.09& 0.695\\
        Regression & image & 0.129 & 0.026 &\textbf{96.21}& 0.646 & 0.165 & 0.032 & \textbf{95.79} & 0.724 \\
        \textbf{Diff4Steer (ours)} & image & \textbf{0.334} & \textbf{0.105} & 89.69& \textbf{0.778} &\textbf{0.341} & \textbf{0.086} & 90.28& \textbf{0.836}\\\midrule
        Regression (txt) & genre &0.378 & 0.103 & \textbf{90.63} & 0.838 & 0.147 & 0.016 &\textbf{92.20} & 0.739\\
        \textbf{Diff4Steer (ours)} & genre & \textbf{0.389} & \textbf{0.108} & 88.02 & \textbf{0.855} & \textbf{0.165} & \textbf{0.019}& 89.65 & \textbf{0.762}\\\midrule
        Regression (txt) & caption & 0.419&\textbf{0.131}&\textbf{90.72}& 0.871 &0.380&\textbf{0.086}&\textbf{91.40}& 0.872\\
        \textbf{Diff4Steer (ours)} & caption & \textbf{0.435} & 0.127 & 87.79& \textbf{0.877} & \textbf{0.384} & 0.085 & 89.67 & \textbf{0.876}\\\midrule
        \textbf{Diff4Steer (ours)} & image + genre & 0.425 & 0.165 & \textbf{91.91} & 0.889 &0.384 & 0.090 & \textbf{94.47} & 0.883\\
        \textbf{Diff4Steer (ours)} & image + caption & \textbf{0.536} & \textbf{0.184}& 91.56 & \textbf{0.915} & \textbf{0.488} & \textbf{0.141} & 93.19 & \textbf{0.916} \\\bottomrule
    \end{tabular}
    }
\end{table}


\begin{table}[t]
    \centering
    \caption{Image-to-music evaluation on MB with genre diversity metrics.}
    \label{tab:i2m-diversity}
    \begin{tabular}{ccccccc}
        \toprule
        $\omega$ & R@10 $\uparrow$ & TA $\uparrow$ & MISCS $\downarrow$ & $\mathcal{H}$@10 $\uparrow$ &$\mathcal{H}$@20 $\uparrow$ & $\mathcal{H}$@50 $\uparrow$ \\
        \midrule
        -1.0 & 0.004 & 0.505 & 0.338 & 1.876 & 2.196 & 2.395  \\
        5.0 &  0.082 & 0.822 & 0.710 & 1.049 & 1.183 & 1.284  \\ 
        9.0 & \textbf{0.089} & 0.832 & 0.772 & 0.920 & 1.030 & 1.113  \\ 
        11.0 &  0.087 & 0.833 & 0.789 & 0.881 & 0.988 & 1.066  \\ 
        15.0 &  0.085 & \textbf{0.834} & 0.807 & 0.843 & 0.945 & 1.019  \\ 
        \bottomrule
    \end{tabular}
\end{table}

\subsection{Retrieval Diversity}
\label{sec:exp-diversity}
\textit{Diff4Steer} generates diverse seed embeddings, as quantified in Table~\ref{tab:i2m-diversity}. For each image, we generate $50$ seed embeddings and measure diversity using MISCS and entropy ($\mathcal{H}$@$K$, with $K \in \{10, 20, 50\}$), calculated on the distribution of ground-truth genres in retrieved music pieces. Varying guided strengths $\omega$ during inference effectively modulates this diversity. Unconditional generation ($\omega=-1.0$) yields the lowest MISCS and highest entropy in recommended genres. Increasing GS initially decreases embedding diversity, with retrieval metrics peaking around $\omega =9.0$ before declining.

\begin{figure}[t]
    \centering
    \includegraphics[width=1.0\linewidth]{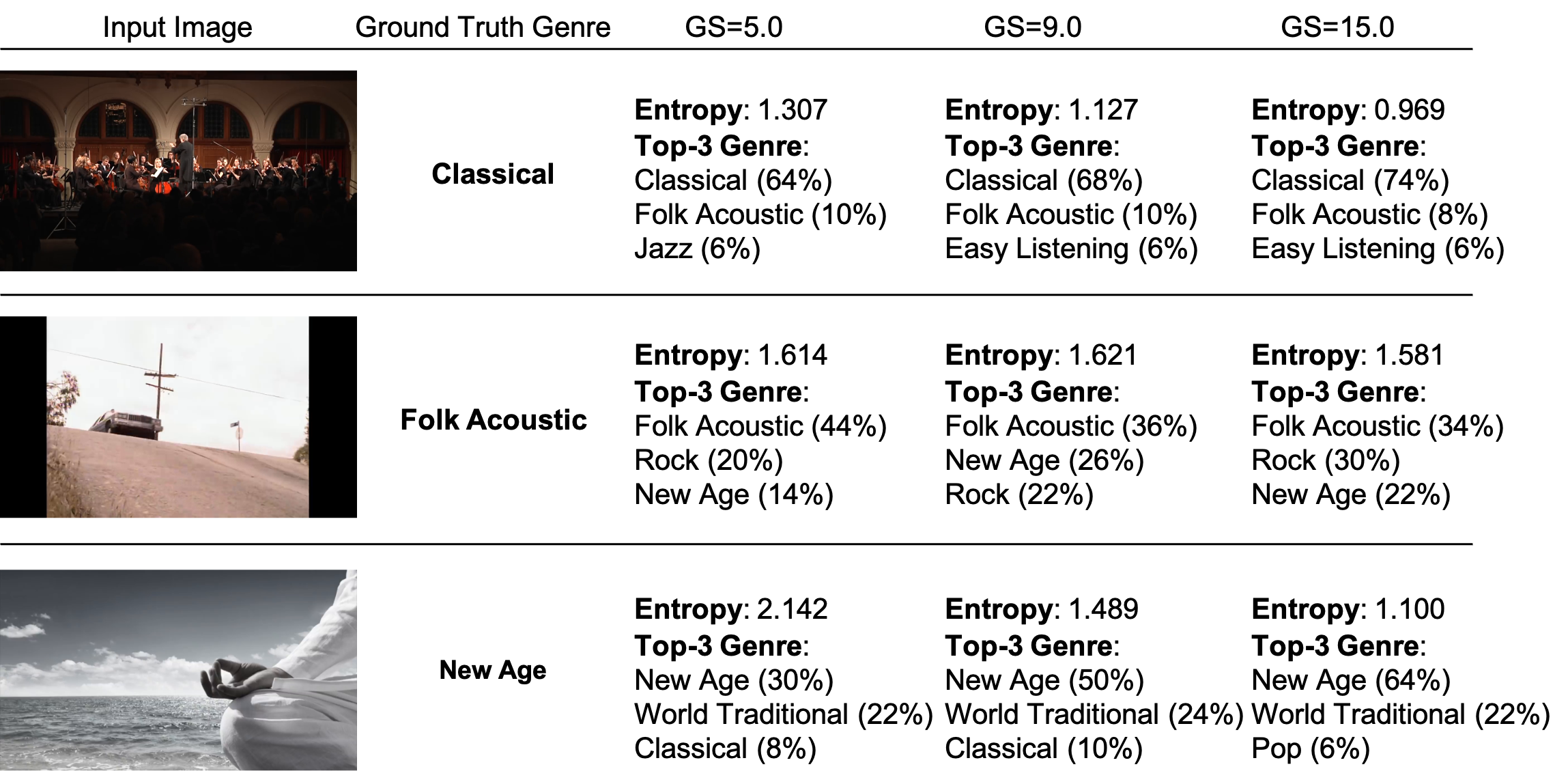}
    \caption{Given an input image and various guided strengths (GS), we generate seed embeddings and retrieve their nearest music piece in MB. We show entropy and the probabilities of Top-3 genres. A higher entropy indicates more diverse music genres of retrieved music pieces.}
    \label{fig:genres}
    \vspace{-0.5em}
\end{figure}

Figure~\ref{fig:genres} illustrates retrieval diversity using three representative input images. With strong image-music correspondence (Top), the entropy is notably lower, reflecting a dominant genre (Classical). Increasing image guidance further amplifies this effect. Conversely, weaker correspondences (Middle, Bottom) show varied entropy changes with increased guidance, sometimes resulting in a dominant genre (Bottom), sometimes maintaining a balance (Middle). In both scenarios, our model generally retrieves music from accurate genres.

\vspace{-0.5em}
\section{Conclusion and limitations}\label{sec:conclusion-limitations}

We introduce a novel generative music retrieval framework featuring a diffusion-based embedding-to-embedding model. By generating non-deterministic seed embeddings from cross-modal queries, our approach improves the quality and diversity of music retrieval results. Our model ensures semantic relevance and high quality, while text-based semantic steering allows user personalization. Extensive evaluations, including personalized retrieval experiments and human studies, show our method's superiority over existing alternatives.





While promising, our framework has limitations as well. High computational demands of diffusion sampling may impede real-time retrieval, and any issues with pre-trained JEMs, such as information loss or underrepresented items, naturally extend to our framework. Additionally, reliance on large, potentially biased training datasets may introduce biases into retrieval results. Future work should address these challenges to improve the retrieval effectiveness of music recommender systems.


\clearpage


\bibliography{reference}

\begin{thebibliography}{21}
\providecommand{\natexlab}[1]{#1}
\providecommand{\url}[1]{\texttt{#1}}
\expandafter\ifx\csname urlstyle\endcsname\relax
  \providecommand{\doi}[1]{doi: #1}\else
  \providecommand{\doi}{doi: \begingroup \urlstyle{rm}\Url}\fi

\bibitem[Abu-El-Haija et~al.(2016)Abu-El-Haija, Kothari, Lee, Natsev, Toderici, Varadarajan, and Vijayanarasimhan]{abu2016yt8m}
S.~Abu-El-Haija, N.~Kothari, J.~Lee, P.~Natsev, G.~Toderici, B.~Varadarajan, and S.~Vijayanarasimhan.
\newblock Youtube-8{M}: A large-scale video classification benchmark.
\newblock \emph{arXiv preprint arXiv:1609.08675}, 2016.

\bibitem[Agostinelli et~al.(2023)Agostinelli, Denk, Borsos, Engel, Verzetti, Caillon, Huang, Jansen, Roberts, Tagliasacchi, et~al.]{agostinelli2023musiclm}
A.~Agostinelli, T.~I. Denk, Z.~Borsos, J.~Engel, M.~Verzetti, A.~Caillon, Q.~Huang, A.~Jansen, A.~Roberts, M.~Tagliasacchi, et~al.
\newblock Musiclm: Generating music from text.
\newblock \emph{arXiv preprint arXiv:2301.11325}, 2023.

\bibitem[Anderson et~al.(2020)Anderson, Maystre, Anderson, Mehrotra, and Lalmas]{anderson2020algorithmic}
A.~Anderson, L.~Maystre, I.~Anderson, R.~Mehrotra, and M.~Lalmas.
\newblock Algorithmic effects on the diversity of consumption on spotify.
\newblock In \emph{Proceedings of the web conference 2020}, pages 2155--2165, 2020.

\bibitem[Chowdhury et~al.(2024)Chowdhury, Nag, KJ, Srinivasan, and Manocha]{chowdhury2024melfusion}
S.~Chowdhury, S.~Nag, J.~KJ, B.~V. Srinivasan, and D.~Manocha.
\newblock Melfusion: Synthesizing music from image and language cues using diffusion models.
\newblock In \emph{Conference on Computer Vision and Pattern Recognition (CVPR)}, 2024.

\bibitem[Covington et~al.(2016)Covington, Adams, and Sargin]{Covington2016DeepNN}
P.~Covington, J.~K. Adams, and E.~Sargin.
\newblock Deep neural networks for youtube recommendations.
\newblock \emph{Proceedings of the 10th ACM Conference on Recommender Systems}, 2016.
\newblock URL \url{https://api.semanticscholar.org/CorpusID:207240067}.

\bibitem[Dhariwal and Nichol(2021)]{dhariwal2021diffusion}
P.~Dhariwal and A.~Nichol.
\newblock Diffusion models beat gans on image synthesis.
\newblock \emph{Advances in neural information processing systems}, 34:\penalty0 8780--8794, 2021.

\bibitem[Elizalde et~al.(2022)Elizalde, Deshmukh, Ismail, and Wang]{elizalde2022clap}
B.~Elizalde, S.~Deshmukh, M.~A. Ismail, and H.~Wang.
\newblock Clap: Learning audio concepts from natural language supervision, 2022.

\bibitem[Heusel et~al.(2017)Heusel, Ramsauer, Unterthiner, Nessler, and Hochreiter]{heusel2017gans}
M.~Heusel, H.~Ramsauer, T.~Unterthiner, B.~Nessler, and S.~Hochreiter.
\newblock Gans trained by a two time-scale update rule converge to a local nash equilibrium.
\newblock \emph{Advances in neural information processing systems}, 30, 2017.

\bibitem[Ho and Salimans(2022)]{ho2022classifier}
J.~Ho and T.~Salimans.
\newblock Classifier-free diffusion guidance.
\newblock \emph{arXiv preprint arXiv:2207.12598}, 2022.

\bibitem[Hoogeboom et~al.(2023)Hoogeboom, Heek, and Salimans]{hoogeboom2023simple}
E.~Hoogeboom, J.~Heek, and T.~Salimans.
\newblock simple diffusion: End-to-end diffusion for high resolution images.
\newblock In \emph{International Conference on Machine Learning}, pages 13213--13232. PMLR, 2023.

\bibitem[Huang et~al.(2022)Huang, Jansen, Lee, Ganti, Li, and Ellis]{huang2022mulan}
Q.~Huang, A.~Jansen, J.~Lee, R.~Ganti, J.~Y. Li, and D.~P. Ellis.
\newblock Mulan: A joint embedding of music audio and natural language.
\newblock \emph{arXiv preprint arXiv:2208.12415}, 2022.

\bibitem[Karras et~al.(2022)Karras, Aittala, Aila, and Laine]{karras2022elucidating}
T.~Karras, M.~Aittala, T.~Aila, and S.~Laine.
\newblock Elucidating the design space of diffusion-based generative models.
\newblock \emph{Advances in Neural Information Processing Systems}, 35:\penalty0 26565--26577, 2022.

\bibitem[Kim et~al.(2007)Kim, Kim, Park, Lee, and Lee]{Kim2007AMR}
D.~M. Kim, K.~Kim, K.-H. Park, J.-H. Lee, and K.-M. Lee.
\newblock A music recommendation system with a dynamic k-means clustering algorithm.
\newblock \emph{Sixth International Conference on Machine Learning and Applications (ICMLA 2007)}, pages 399--403, 2007.
\newblock URL \url{https://api.semanticscholar.org/CorpusID:17603549}.

\bibitem[Kingma and Ba(2014)]{kingma2014adam}
D.~P. Kingma and J.~Ba.
\newblock Adam: A method for stochastic optimization.
\newblock \emph{arXiv preprint arXiv:1412.6980}, 2014.

\bibitem[Koren et~al.(2009)Koren, Bell, and Volinsky]{Koren2009MatrixFT}
Y.~Koren, R.~M. Bell, and C.~Volinsky.
\newblock Matrix factorization techniques for recommender systems.
\newblock \emph{Computer}, 42, 2009.
\newblock URL \url{https://api.semanticscholar.org/CorpusID:58370896}.

\bibitem[Liu et~al.(2023)Liu, Tian, Yuan, Liu, Mei, Kong, Wang, Wang, Wang, and Plumbley]{liu2023audioldm}
H.~Liu, Q.~Tian, Y.~Yuan, X.~Liu, X.~Mei, Q.~Kong, Y.~Wang, W.~Wang, Y.~Wang, and M.~D. Plumbley.
\newblock Audioldm 2: Learning holistic audio generation with self-supervised pretraining.
\newblock \emph{arXiv preprint arXiv:2308.05734}, 2023.

\bibitem[Loshchilov and Hutter(2016)]{loshchilov2016sgdr}
I.~Loshchilov and F.~Hutter.
\newblock Sgdr: Stochastic gradient descent with warm restarts.
\newblock In \emph{International Conference on Learning Representations}, 2016.

\bibitem[Radford et~al.(2021)Radford, Kim, Hallacy, Ramesh, Goh, Agarwal, Sastry, Askell, Mishkin, Clark, et~al.]{radford2021learning}
A.~Radford, J.~W. Kim, C.~Hallacy, A.~Ramesh, G.~Goh, S.~Agarwal, G.~Sastry, A.~Askell, P.~Mishkin, J.~Clark, et~al.
\newblock Learning transferable visual models from natural language supervision.
\newblock In \emph{International conference on machine learning}, pages 8748--8763. PMLR, 2021.

\bibitem[Rombach et~al.(2022)Rombach, Blattmann, Lorenz, Esser, and Ommer]{rombach2022high}
R.~Rombach, A.~Blattmann, D.~Lorenz, P.~Esser, and B.~Ommer.
\newblock High-resolution image synthesis with latent diffusion models.
\newblock In \emph{Proceedings of the IEEE/CVF conference on computer vision and pattern recognition}, pages 10684--10695, 2022.

\bibitem[Schneider et~al.(2023)Schneider, Kamal, Jin, and Sch{\"o}lkopf]{schneider2023mo}
F.~Schneider, O.~Kamal, Z.~Jin, and B.~Sch{\"o}lkopf.
\newblock Mo$\backslash$\^{} usai: Text-to-music generation with long-context latent diffusion.
\newblock \emph{arXiv preprint arXiv:2301.11757}, 2023.

\bibitem[Song et~al.(2020)Song, Sohl-Dickstein, Kingma, Kumar, Ermon, and Poole]{song2020score}
Y.~Song, J.~Sohl-Dickstein, D.~P. Kingma, A.~Kumar, S.~Ermon, and B.~Poole.
\newblock Score-based generative modeling through stochastic differential equations.
\newblock In \emph{International Conference on Learning Representations}, 2020.

\end{thebibliography}
\bibliographystyle{abbrvnat}
\setcitestyle{numbers} 

\clearpage
\appendix
\section{Experiment details}\label{sec:appendix-exp-details}

\subsection{Training and sampling}
\label{sec:appendix-training}
We provide additional details for training and sampling, including the architecture of the backbone model, diffusion training \& sampling details, and the computational efficiency of our Diff4Steer model.

\subsubsection{Architecture}
\begin{figure}[ht]
    \centering
    \includegraphics[width=0.3\linewidth]{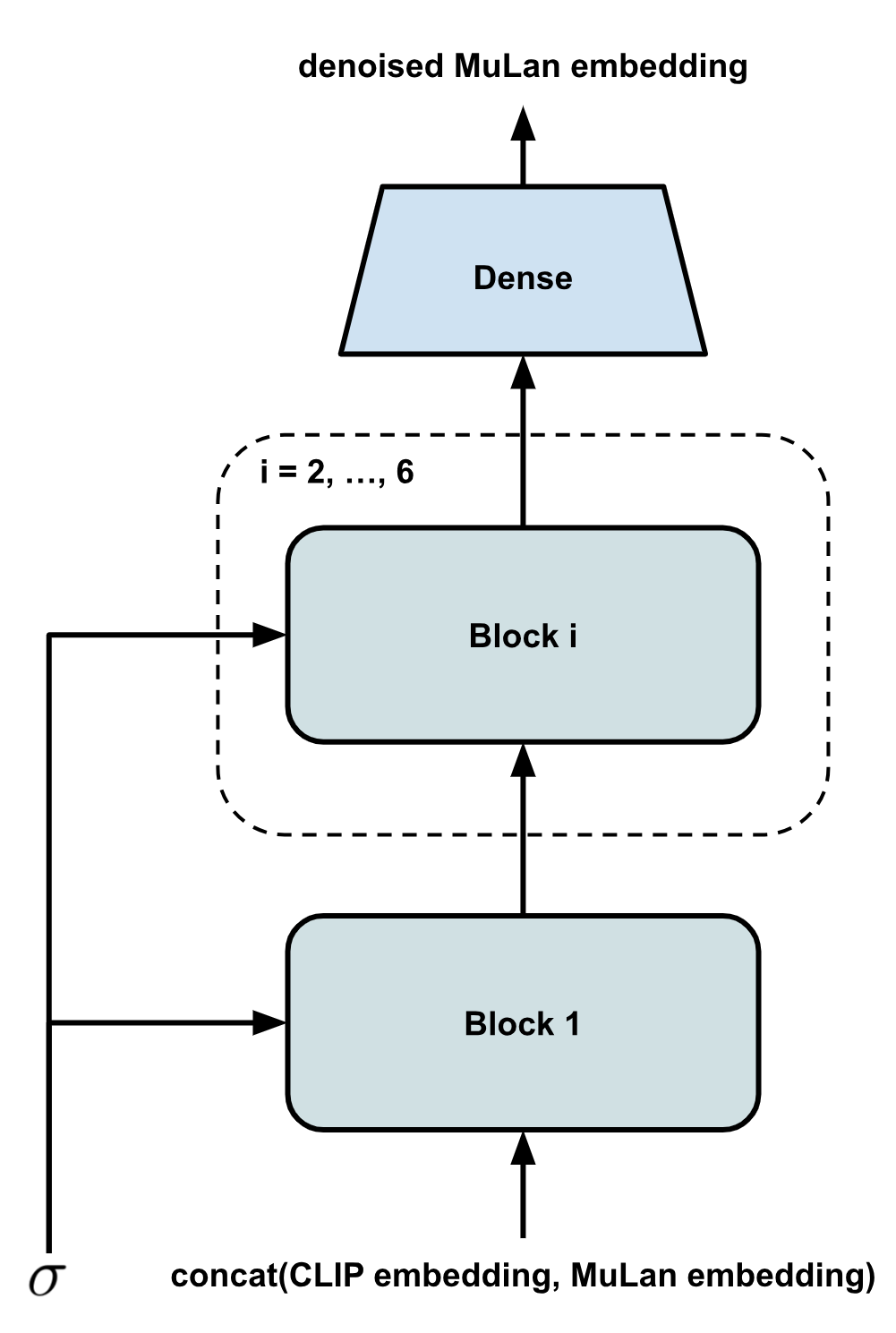}
    \caption{Overall architecture for the diffusion backbone.}
    \label{fig:architecture-overall}
\end{figure}

\begin{figure}[ht]
    \centering
    \includegraphics[width=0.48\linewidth]{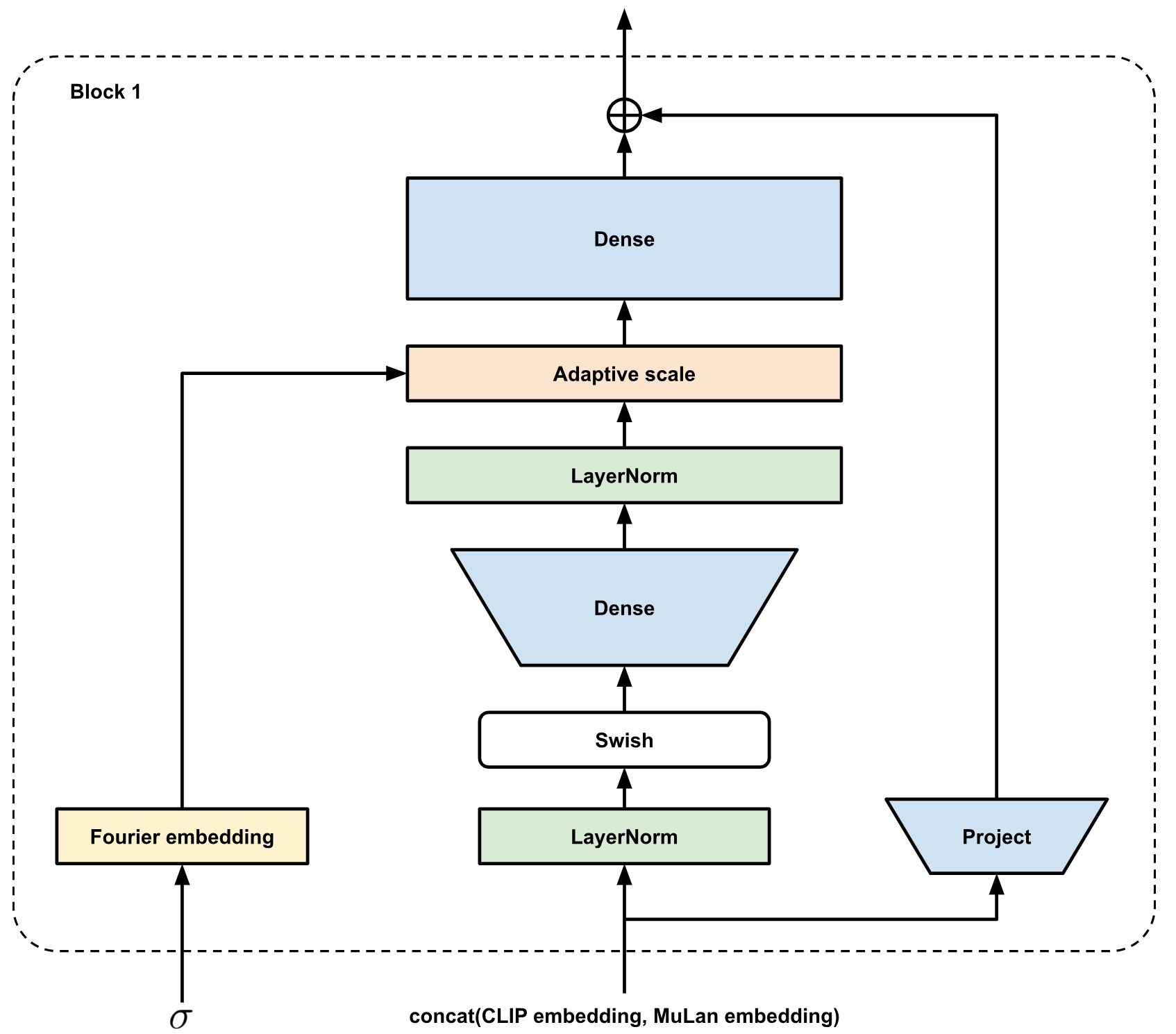}
    \includegraphics[width=0.48\linewidth]{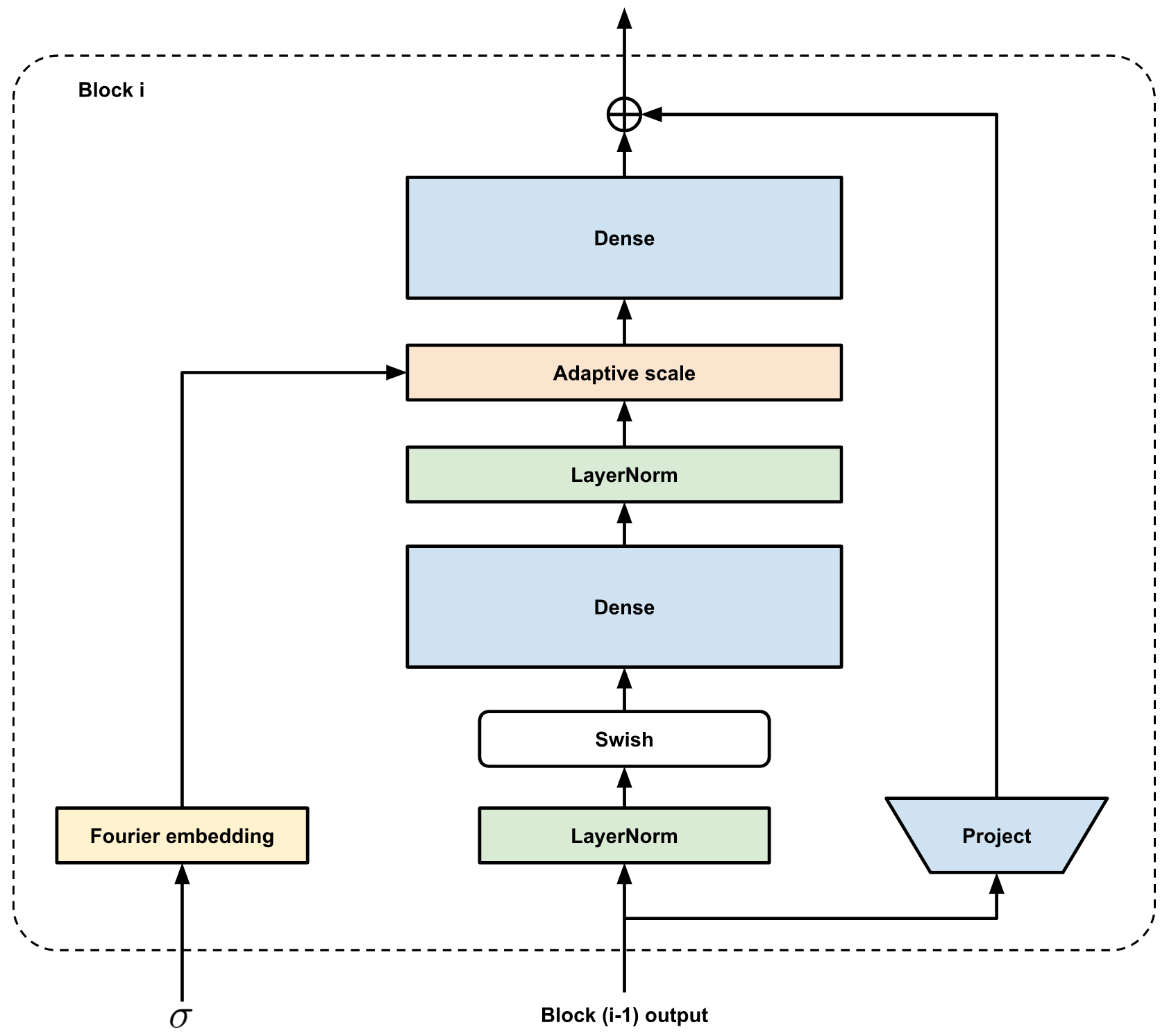}
    \caption{Architecture diagram for the ResNet blocks.}
    \label{fig:architecture-blocks}
\end{figure}

Our diffusion backbone model is a ResNet model consisting of 6 ResNet blocks, followed by a final linear projection layer. We incorporate the noise level by adding an adaptive scaling layer similar to~\citet{dhariwal2021diffusion}. The overall architecture is shown in Figure~\ref{fig:architecture-overall}, and detailed architecture of each ResNet block is shown in Figure~\ref{fig:architecture-blocks}.

\subsubsection{Diffusion model}

\textbf{Noise schedule} $\sigma_t := \sigma(t)$. 
We employ a variance exploding (VE) diffusion scheme~\citep{song2020score}
\begin{align}
    \sigma(t) = \sigma_{\text{data}}\cdot\hat{\sigma}(t),
\end{align}
where $t\in[0, 1]$ and $\sigma_\text{data}$ is the population standard deviation of the samples (for MuLan music embeddings we have $\sigma_\text{data} \approx 0.088$). $\hat{\sigma}(t)$ follows a tangent noise schedule:
\begin{align}
    \hat{\sigma}(t) = \sigma_{\max} \cdot \frac{\tan (\alpha_{\max}t)}{\tan(\alpha_{\max})},
\end{align}
where $\alpha_{\max} = 1.5$ and $\sigma_{\max} = 100.0$. This schedule is in essence linearly re-scaling the $\tan(\cdot)$ function in domain $[0, \alpha_{\max}]$ to fall within range $[0, \sigma_{\max}]$. It is similar to the shifted cosine schedule proposed in~\citep{hoogeboom2023simple} (note that a tangent schedule in $\sigma$ is equivalent to a cosine schedule in the $1/(\sigma^2 + 1)$).

The forward diffusion process ($t$ goes from 0 to 1) follows the SDE
\begin{align}
    d\tilde{z}_{\text{m},t} =  \sqrt{2\dot{\sigma}_t\sigma_t}\;dW_t,
\end{align}
whose marginal distributions $p(\tilde{z}_{\text{m},t})$ matches those of the reverse-time ($t$ goes from 1 to 0) sampling SDE~\eqref{eq:sampling_sde} for all $t$.

\textbf{Preconditioning. } 
In order to train more efficiently, we apply preconditioned parametrization for the denoising model:
\begin{align}
    D_\theta(\tildezm, \sigma, q) = c_\text{skip}(\sigma) \tildezm + c_\text{out}(\sigma) F_\theta (c_\text{in}(\sigma) \tildezm, c_\text{noise}(\sigma) , q),
\end{align}
where $F_\theta$ is the raw neural network backbone function. 
The rationale is to have approximately standardized input and output distributions for $F_\theta$.
Following~\citet{karras2022elucidating}, we set the preconditioning coefficients to be:
\begin{align}
    &c_\text{skip}(\sigma) = \sigma_\text{data}^2 / (\sigma^2 + \sigma_\text{data}^2)\\
    &c_\text{out}(\sigma) = \sigma \cdot \sigma_\text{data} / \sqrt{\sigma_\text{data}^2 + \sigma^2}\\
    &c_\text{in}(\sigma) = 1 / \sqrt{\sigma^2 + \sigma_\text{data}^2} \\
    &c_\text{noise}(\sigma) = \frac{1}{4} \log (\sigma).
\end{align}

\textbf{Noise sampling} $\eta(\sigma)$.
During training, we use the following noise sampling scheme:
\begin{align}
\label{eq:noise_sampling}
    \sigma = \sigma_{\min} \left(\frac{\sigma_\text{data} \sigma_{\max}}{\sigma_{\min}}\right)^\delta,
\end{align}
where $\delta \sim \mathcal{U}[0, 1]$. $\sigma_{\min} = 10^{-4}$ is a clip value for the minimum noise level to prevent numerical blow-up. For each training example $\zm$, we use \eqref{eq:noise_sampling} to sample a level $\sigma$ and subsequently a noisy input $\tilde{z}_\text{m} = \zm + \sigma\epsilon$.

\textbf{Noise weighting} $\lambda(\sigma)$.
We use the EDM weighting~\citep{karras2022elucidating}:
\begin{align}
    \lambda(\sigma) = \frac{\sigma_\text{data}^2 + \sigma^2}{\sigma_\text{data} \sigma},
\end{align}
which is designed to have an effective weighting that is uniform across all noise levels.

\textbf{Solver time schedule. } 
We adopt the solver time schedule in~\citet{karras2022elucidating} with $\rho=7$, and use $N=256$ sampling steps in our experiments.
\begin{align}
    t_{i<N} = \bigg(\sigma_{\max}^{1/\rho} + \frac{i}{N-1}(\sigma_{\min}^{1/\rho} - \sigma_{\max}^{1/\rho})\bigg)^\rho.
\end{align}

\subsubsection{Computational efficiency}
\label{sec:appendix-compute}
It is efficient to run sampling with our lightweight model, which can be fit into one TPU device. With JIT compilation in JAX, sampling 4 seed embeddings with one query embedding takes around 0.8 ms. Batch sampling for 4,000 query embeddings takes around 3.3 ms.

\subsection{Optimization and hyperparameter search}
For Adam optimizer, we use a peak learning rate of $10^{-5}$ for 2,000,000 steps, with 10,000 warm-up steps and initial learning rate 0. We searched over learning rates 1e-5, 3e-5, 1e-4, 3e-4, 1e-3, and conditional mask probabilities 0.1 and 0.3.


\subsection{Details on Human Study}\label{sec:appendix-human-study}
\begin{figure}[h]
    \centering
    \includegraphics[width=\linewidth]{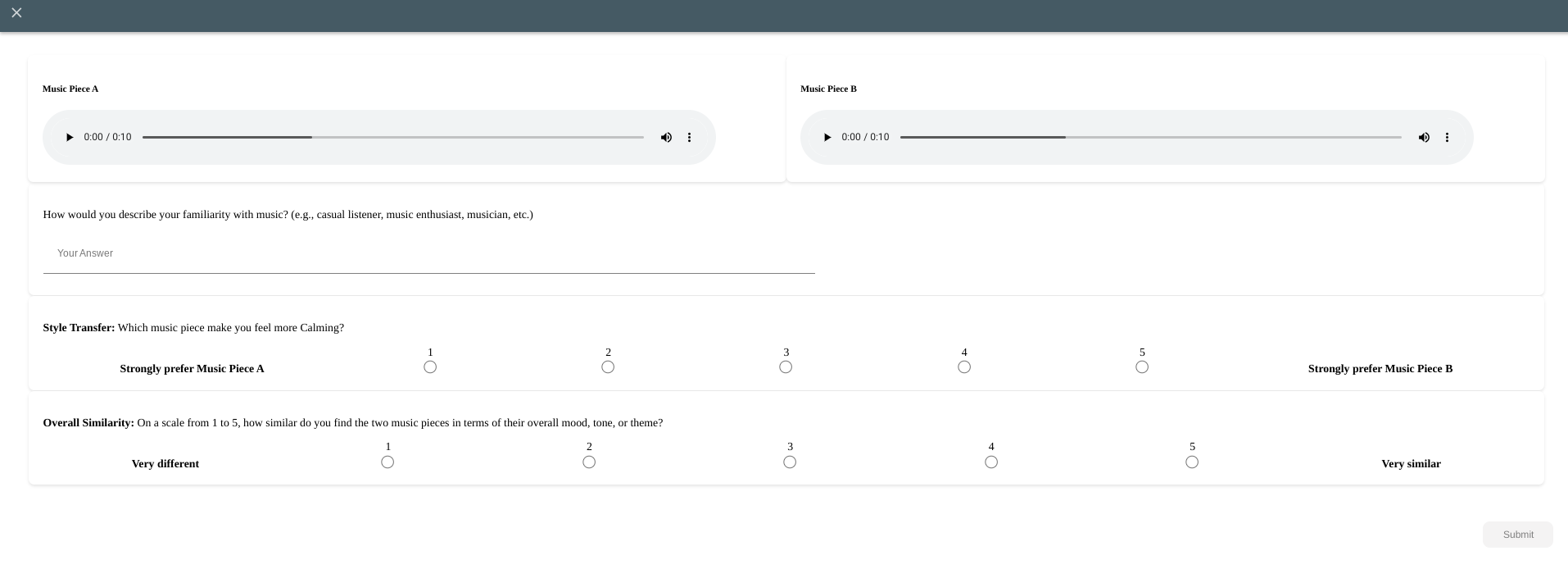}
    \caption{An example questionnaire used for human study.}
    \label{fig:questionnaire}
\end{figure}

In human study, the participants are asked to listen to two music clips, and rate on a scale of 1 to 5:
\begin{enumerate}
    \item Which music piece make you fiel more $\langle$ insert music style $\rangle$?
    \item How similar do you find the two music pieces in terms of their overall mood, tone, or theme?
\end{enumerate}
Screenshot of an example questionnaire is shown in Figure~\ref{fig:questionnaire}.

The human study result is reported for a fixed range of steering strength, between $0.06, 0.08$ and interpolation ratio of $0.55$.

\section{Detailed metric definitions}
\label{sec:appendix-metrics}
We provide detailed metric definitions in this section.
\subsection{Fr\'{e}chet MuLan Distance (FMD)}
Given a set of generated MuLan embeddings $\mathcal{S}_\text{G} \subset \Zm$ and a set of real MuLan embeddings $\mathcal{S}_\text{R} \subset \Zm$, and let $\mu_\text{G}$, $\mu_\text{R}$ and $\Sigma_\text{G}$, $\Sigma_\text{R}$ be the mean and variance of $\mathcal{S}_\text{G}$ and $\mathcal{S}_\text{R}$ respectively. The Fr\'{e}chet MuLan Distance is defined as
\begin{align*}
    \text{FMD} :=  \Vert \mu_\text{G} - \mu_\text{R} \Vert^2 + \text{tr}\big(\Sigma_\text{G} + \Sigma_\text{R} - 2 (\Sigma_\text{G}\Sigma_\text{R})^{1/2}\big).
\end{align*}
To compute FMD, we use a large dataset containing around 63M MuLan audio embeddings as the set of real embeddings and estimate $\mu_R, \Sigma_R$.

\subsection{Mean intra-sample cosine similarity (MISCS)} 
Given $n$ generated MuLan embeddings $\zm^{(1)}, \dots, \zm^{(n)}$ (rescaled to norm 1), the mean intra-sample cosine similarity is defined as
\begin{align*}
    \cos (\zm^{(1)}, \dots, \zm^{(n)}) := \frac{2}{n(n-1)}\sum_{i=1}^n \sum_{j=1}^{i-1} \langle \zm^{(i)}, \zm^{(j)}\rangle.
\end{align*}
\label{diversity}

\subsection{Cross-modal alignment}

\textbf{Music-image alignment (M2I).} 
Given a set of images $\mathcal{D}_\text{i}$ and a separate set of texts $\mathcal{D}_\text{txt}$, M2I is defined as the average of the product of image-text and text-music similarities:
\begin{align*}
        \text{M2I} := \frac{1}{\vert\mathcal{D}_\text{i}\vert} \sum_{\zi \in \mathcal{D}}\sum_{\zt \in \mathcal{D}_\text{txt}} \langle \zi, \zt^\text{CLIP}\rangle \cdot \langle \zm^\text{pred} (\zi), \zt^{\text{MuLan}}\rangle.
\end{align*}

In addition to M2I, similar metrics can be defined to measure music-to-music (M2M) and music-to-caption alignments (M2C).

\textbf{Music-music alignment (M2M) }
Given a dataset $\mathcal{D}_\text{M2M}$ that contains paired image and music data, M2M is the average dot product between the predicted MuLan embeddings $\zm^\text{pred} (\zi)$ and the ground truth MuLan embeddings.\footnote{There is a slight abuse of notations, as the prediction might not be deterministic. For diffusion model, $\zm^\text{pred}$ is a sample of the prediction.}
\begin{align*}
    \text{M2M} := \frac{1}{\vert\mathcal{D}_\text{M2M}\vert}\sum_{\zi, \zm \in \mathcal{D}_\text{M2M}} \langle \zm^\text{pred} (\zi), \zm\rangle
\end{align*}

\textbf{Music-caption alignment (M2C) } 
Given a dataset $\mathcal{D}_\text{M2C}$ that contains paired image and music caption data, M2C is the average dot product between the predicted MuLan embeddings $\zm^\text{pred} (\zi)$ and the MuLan embeddings for music captions.
\begin{align*}
    \text{M2C} := \frac{1}{\vert\mathcal{D}_\text{M2C}\vert}\sum_{\zi, z_\text{cap} \in \mathcal{D}_\text{M2C}} \langle\zm^\text{pred} (\zi), z_\text{cap}\rangle
\end{align*}

\subsection{Triplet Accuracy (TA)}
Given a dataset containing triplets $\mathcal{D}_t = \{(z_{\text{anchor}}), z_{+}, z_{-})\}$, where $z_\text{anchor}$ is the image or text input, $z_{+}$ and $z_{-}$ are positive and negative examples respectively. Let $z_\text{pred}$ be a sample of the model prediction, the triplet accuracy is defined as,
\begin{align*}
    \text{Triplet accuracy} := \mathbb{E}_{(z_{\text{anchor}}, z_{+}, z_{-}) \sim \mathcal{D}_t} \mathbbm{1}[\langle z_\text{pred}, z_{+}\rangle \geq \langle z_\text{pred}, z_{-}\rangle].
\end{align*}

\subsection{Entropy}
The entropy $\mathcal{H}$ (@K) is computed as
\begin{align*}
    \mathcal{H} := -\sum_l^{L} p_l\log(p_l),\;\; p_l = n_l / K
\end{align*}
where $L$ denotes the total number of genres. $n_l$ is the count in genre $l$ amongst $K$ predictions/samples.

\subsection{Recall@K}
The Recall@K metric is defined as
\begin{align*}
    \text{R@K} := \frac{\text{number of retrieved items @K that are relevant}}{\text{total number of relevant items}}.
\end{align*}

\subsection{Human Study Evaluation Metrics}
\textit{Relevance (REL) }
To evaluate the relevance of steered music to a given semantic concept, users are asked to compare the mood or style of a reference music piece to the steered piece on a 5-point scale. A score of 4 or 5 is considered a win for positive steering, while a score of 1 or 2 indicates success in negative steering.

\textit{Consistency (CON) }
To assess the consistency of the overall theme and tone in steered music, users compare it to a reference piece, rating their similarity on a 5-point scale. This score is then mapped to a 0-100 range to reflect the degree of consistency.

\section{Multi-modal prompts for music caption generation}
\label{sec:appendix-prompt}

We use interleaved prompts of image and music captions with the multi-modal Gemini model. Some examples are shown in Table~\ref{tab:prompts}.

\begin{table}[ht]
    \centering
    \caption{Interleaved Multi-modal prompts for Gemini Baseline. For a given image the prompt will generate image caption and music caption.}
    \label{tab:prompts}
    \begin{tabular}{|m{4cm}|m{9cm}|}
        \hline
        \textbf{Image} & \textbf{Description} \\ \hline
        \includegraphics[width=4cm]{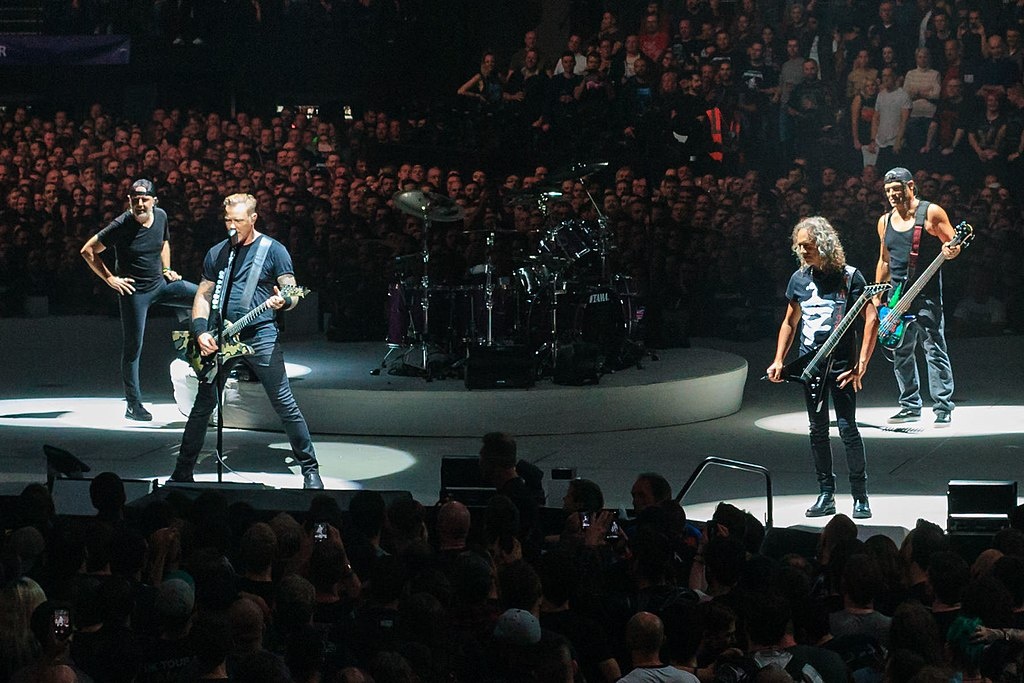} & [Image caption] Dynamic energy of a live concert, band members engaging with the crowd, instruments at the ready. The stage glows under the spotlight, and the audience is a sea of faces, the air charged with anticipation and excitement. 
        
        [Music caption] Electrifying guitar riff, powerful drums and bass line. \\ \hline
        \includegraphics[width=4cm]{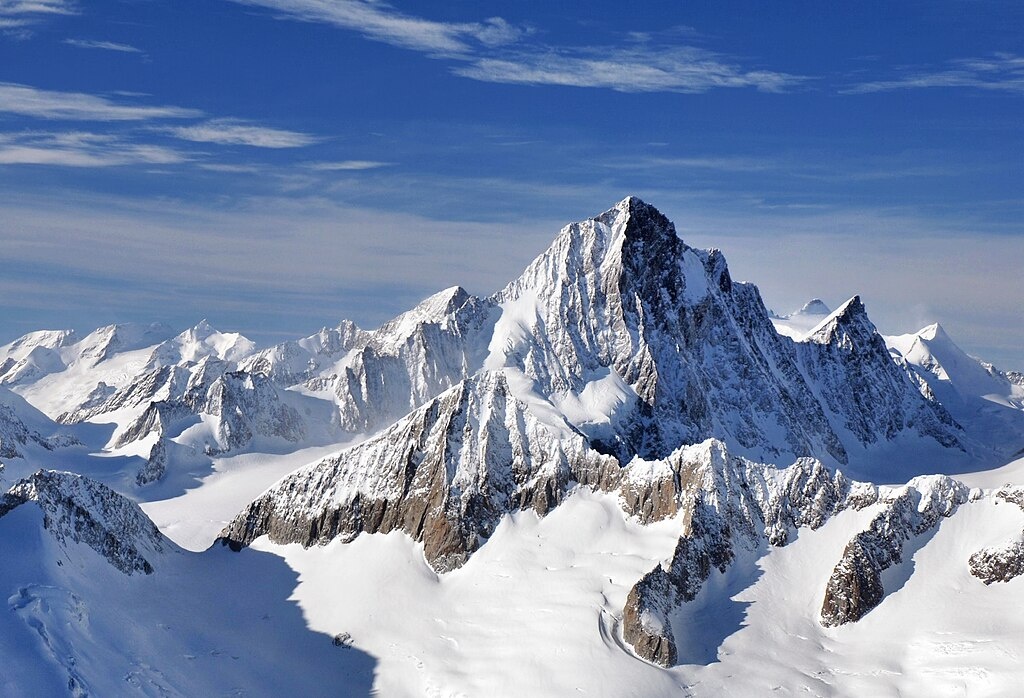} & [Image caption] Majestic grandeur of a snow-capped mountain range under a clear blue sky. The rugged peaks rise sharply, their jagged lines softened by blankets of pristine snow. Shadows and light play across the slopes, creating a tapestry of blue and white that speaks to the silent power of nature.
        
        [Music caption] Orchestral score with string section. \\ \hline
        \includegraphics[width=4cm]{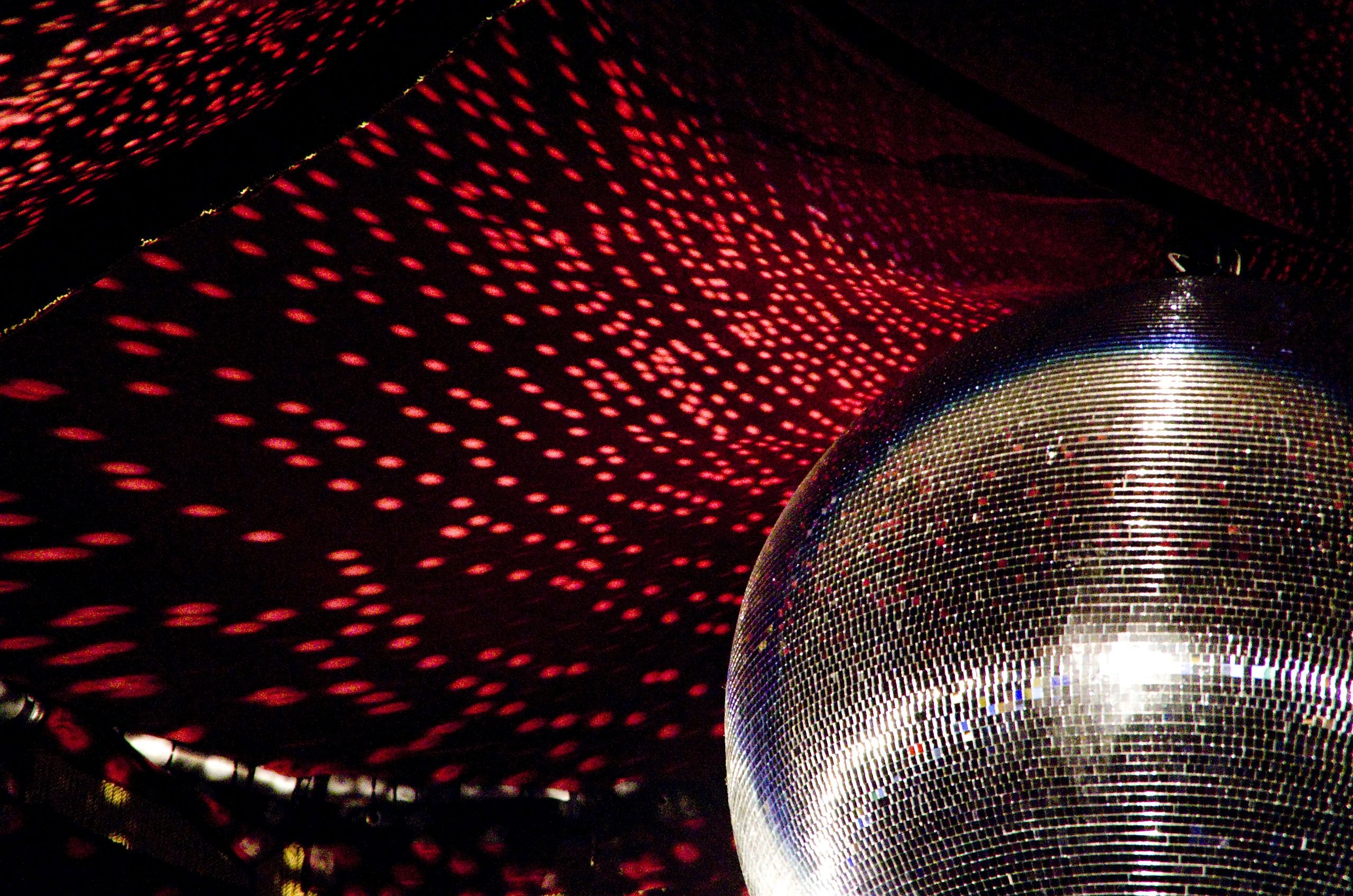} & [Image caption] Dynamic interplay of light and reflection as a disco ball casts a constellation of red dots across a dark expanse. The mirrored surface fragments the light, creating a pattern that suggests both the energy of a dance floor and the cosmic expanse of a starlit sky. 
        
        [Music caption] Rhythmic, upbeat electronic dance track. \\ \hline
        \includegraphics[width=4cm]{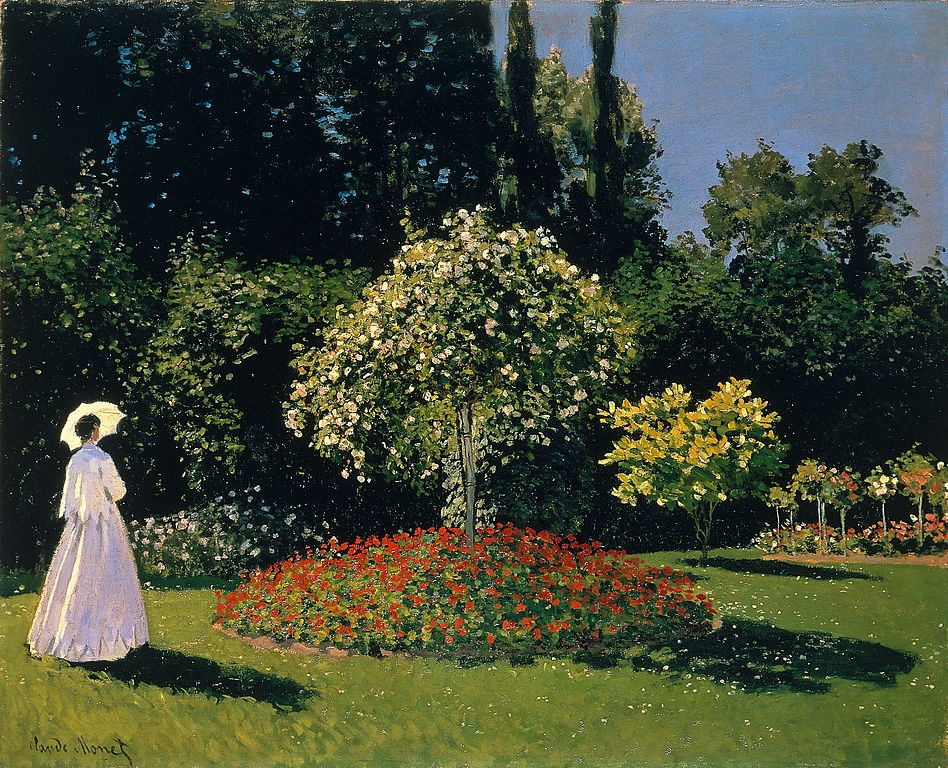} & [Image caption] A solitary figure stands enveloped by the quietude of a vibrant garden, basking in the gentle embrace of sunlight. She seems to be in a moment of tranquil reflection, as the world around her bursts with the life of untamed blooms and the soft whisper of leaves in the breeze. It is an image of peaceful solitude, where the clamor of the world falls away before the simple purity of nature's own artistry.
        
        [Music caption] Soft piano melody with a gentle cello. \\ \hline
        \includegraphics[width=4cm]{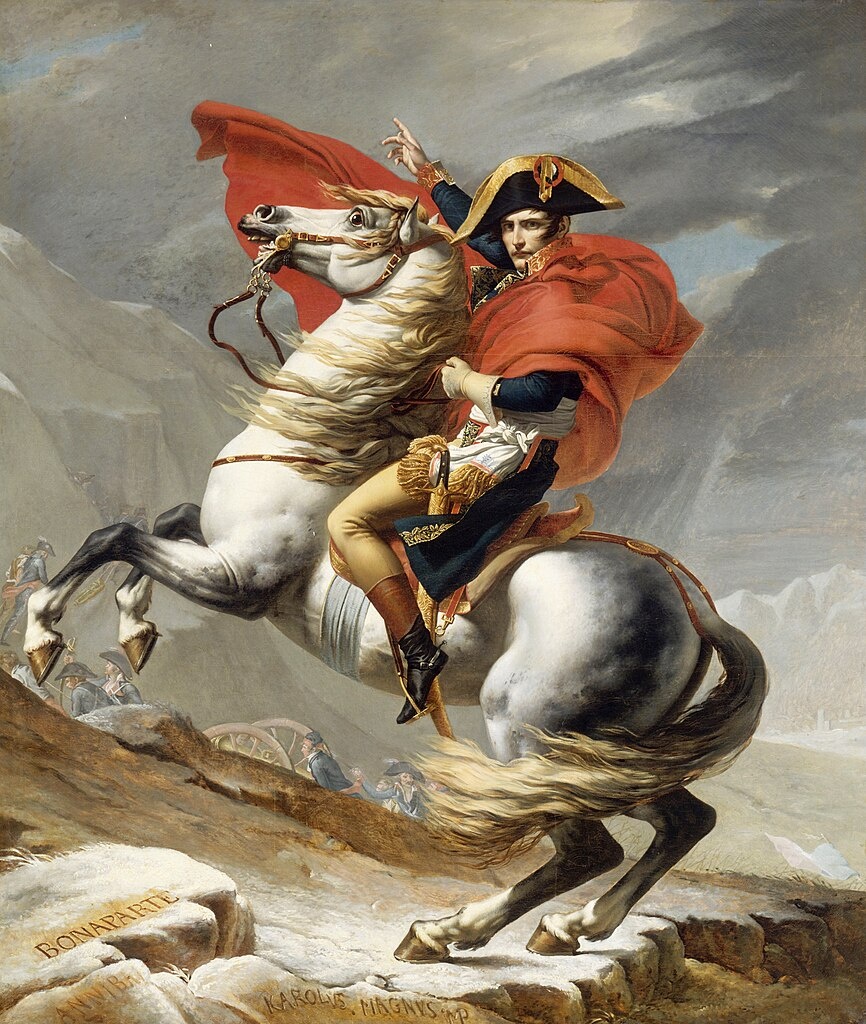} & [Image caption] Commanding figure on horseback, steeped in the iconography of power and leadership. The rider, cloaked in a flowing cape, gestures assertively forward, a symbol of bold ambition. The rearing horse adds to the dramatic intensity, with muscular detail and a mane tossed by the vigor of movement. The backdrop is sparse, the sky subdued, focusing the viewer’s attention on the figure that dominates the scene, a representation of determination and the forging of destiny.
        
        [Music caption] Epic battle atmosphere, full orchestra, strong brass section and rolling drums. \\ \hline
    \end{tabular}
\end{table}
\section{More Alignment Analysis}
\label{sec:appendix-alignment}

Table~\ref{tab:qual-align-musiccaps} reports image-to-music alignment results on the extended MusicCaps and the MelBench datasets. For the deterministic baseline and our approach, the output is audio embedding. The ``reference'' is the alignment metric values with the ground truth MuLan audio embeddings. While for the Gemini baselines, the output is text embedding. The ``reference'' is the alignment metric values with the ground truth MuLan text embeddings. 

Compared to the Gemini baseline, our approach gets higher music-music and music-image alignment score. While the Gemini image caption + MuLan baseline has better music-caption alignment score. To our surprise, the deterministic baseline gets highest alignment score on the MelBench dataset, suggesting that images from MelBench has similar distribution over our training data YT8M. However, according to the FMD quality metrics, the deterministic baseline yield low quality embedding. 
\begin{table}[t]
    \centering
    \caption{Embedding alignment with ground truth image and text, evaluated on the MusicCaps and MelBench dataset.}
    \label{tab:qual-align-musiccaps}
    \begin{tabular}{ccccccc}\toprule
        & \multicolumn{2}{c}{M2M $\downarrow$} & \multicolumn{2}{c}{M2C $\downarrow$} & \multicolumn{2}{c}{M2I $\downarrow$} \\
        & MC & MB & MC & MB & MC & MB \\\midrule
        Reference (gt text) & 47.41 & 47.36 & 100 & 100 & 87.83 & 91.26\\
        Gemini (img cap.) & 28.13 &34.46 & 28.25 &36.64 & 89.12 & 90.32 \\ 
        Gemini (music cap.) & 20.63 &32.13 & 21.30 &35.85 & 84.48 & 88.09\\\midrule 
        Reference (gt audio) & 100 & 100 & 47.41 & 47.39 & 90.77 & 88.59\\
        Deterministic & 45.97 & \textbf{50.64} & 41.77 & \textbf{45.47} & 96.21 & \textbf{95.79} \\ 
        Ours (w/o data aug.) & 40.28 & 44.35 & 33.29 & 36.48 & 91.60 & 91.06 \\ 
        \bottomrule
    \end{tabular}
\end{table}

\section{Additional evaluation results}\label{sec:appendix-additional-results}
\begin{figure}[ht]
    \centering
    \includegraphics[width=0.48\linewidth]{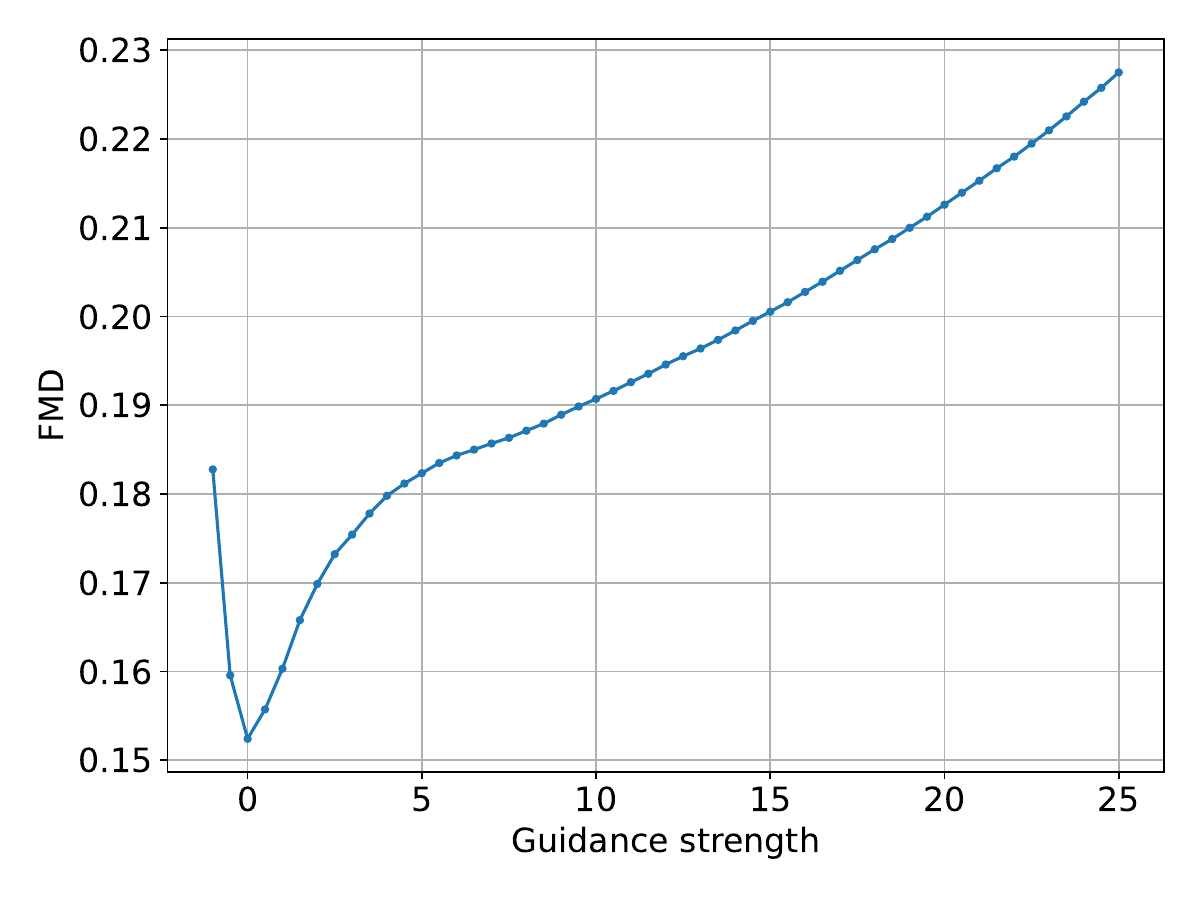}
    \includegraphics[width=0.48\linewidth]{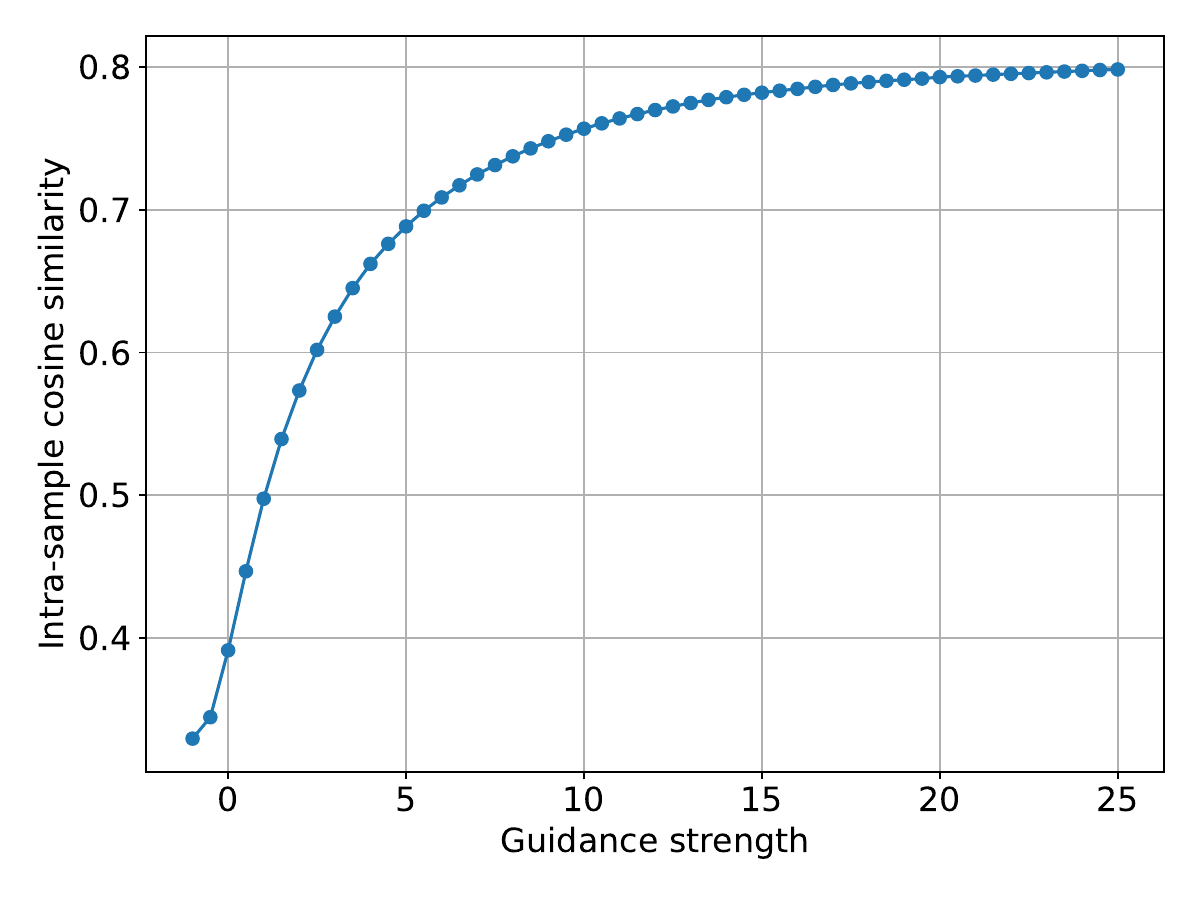}
    \caption{FMD (left) and MISCS (right) as functions of image CFG strength $\omega$. Metrics are computed for the image-to-music task evaluated on the MusicCaps dataset.}
    \label{fig:fmd-gs-musiccaps}
\end{figure}

\begin{figure}[ht]
    \centering
    \includegraphics[width=0.48\linewidth]{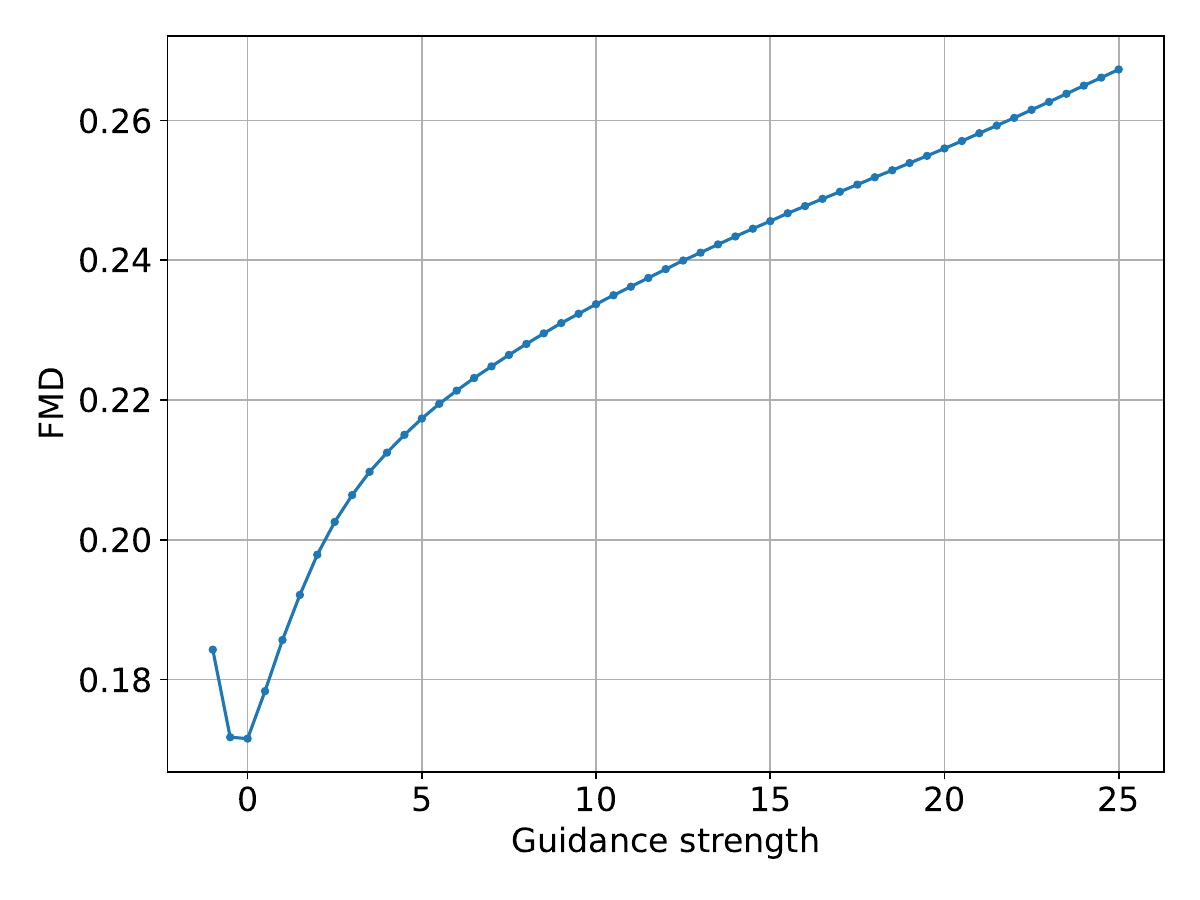}
    \includegraphics[width=0.48\linewidth]{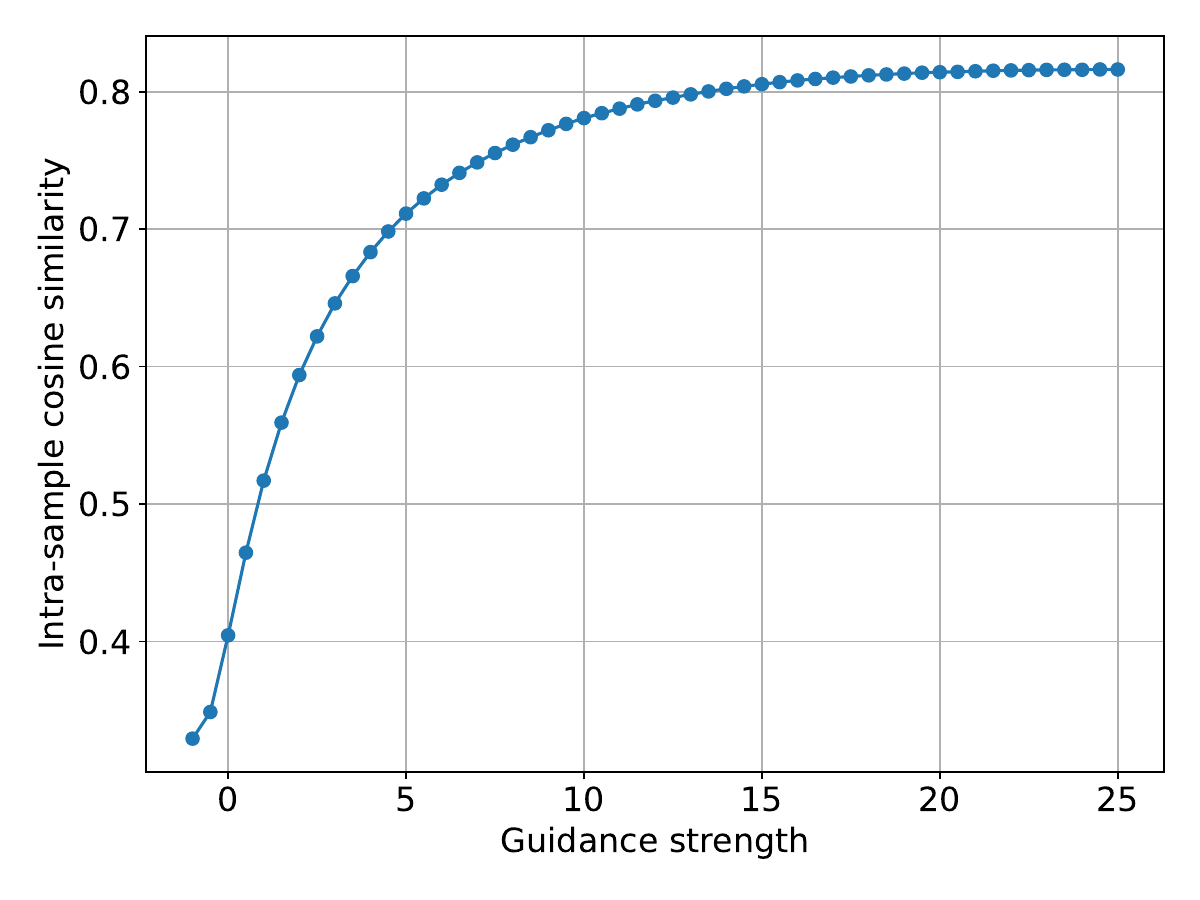}
    \caption{FMD (left) and MISCS (right) as functions of image CFG strength $\omega$. Metrics are computed for the image-to-music task evaluated on the MelBench dataset.}
    \label{fig:fmd-gs-melbench}
\end{figure}

\begin{figure}[ht]
    \centering
    \includegraphics[width=0.48\linewidth]{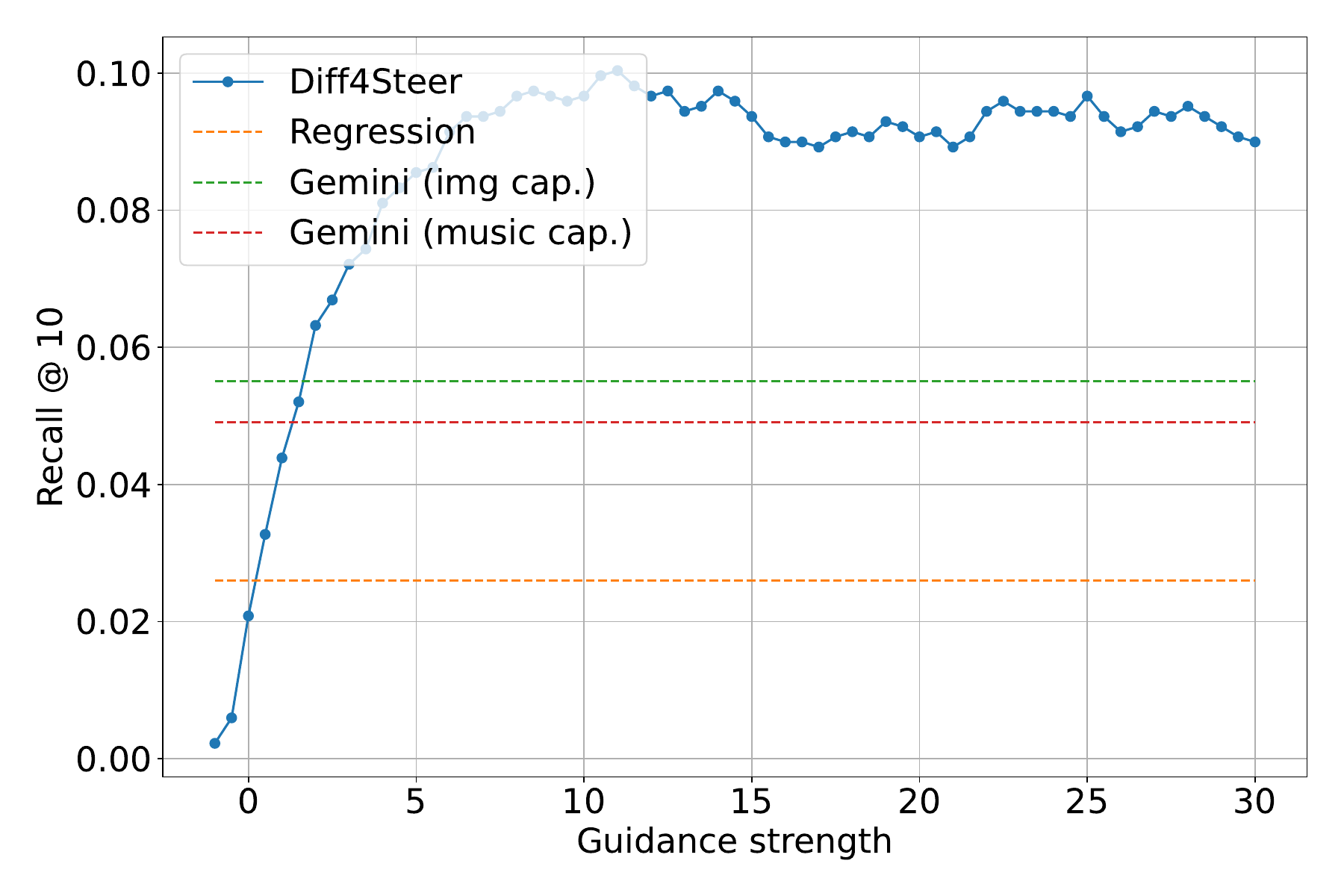}
    \includegraphics[width=0.48\linewidth]{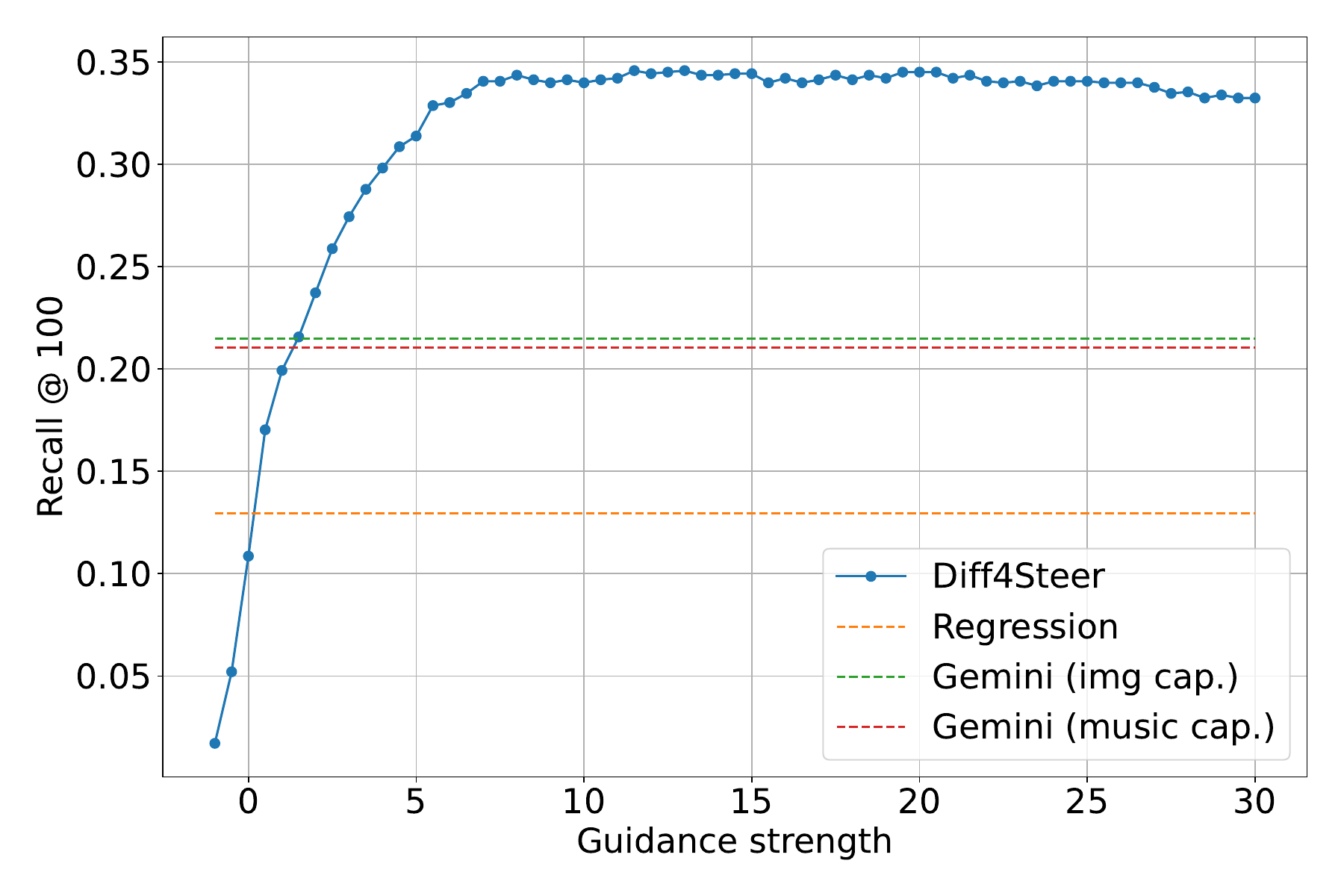}
    \caption{Recall@10 (left) and Recall@100 (right) vs. image CFG strengths on the MusicCaps dataset.}
    \label{fig:recall-vs-gs-musiccaps}
\end{figure}

\begin{figure}[ht]
    \centering
    \includegraphics[width=0.48\linewidth]{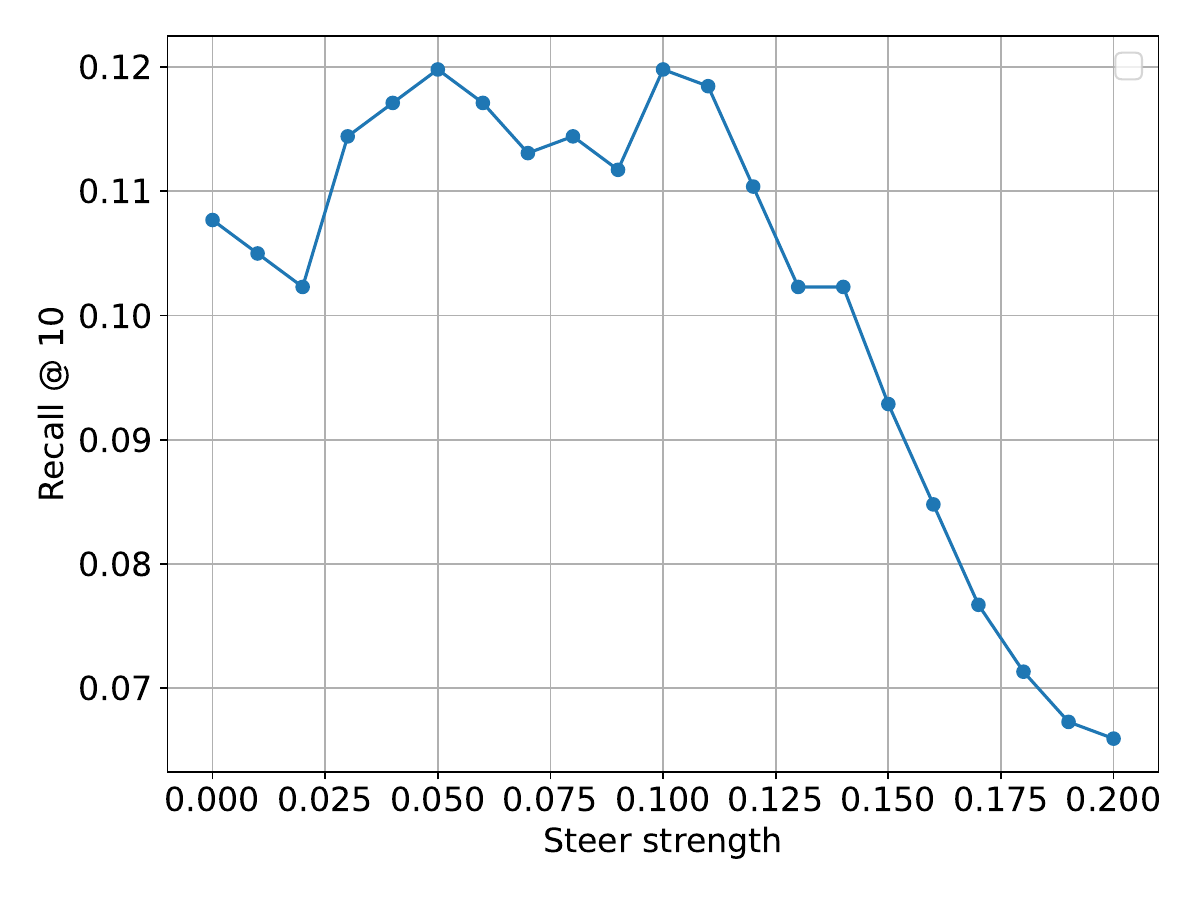}
    \includegraphics[width=0.48\linewidth]{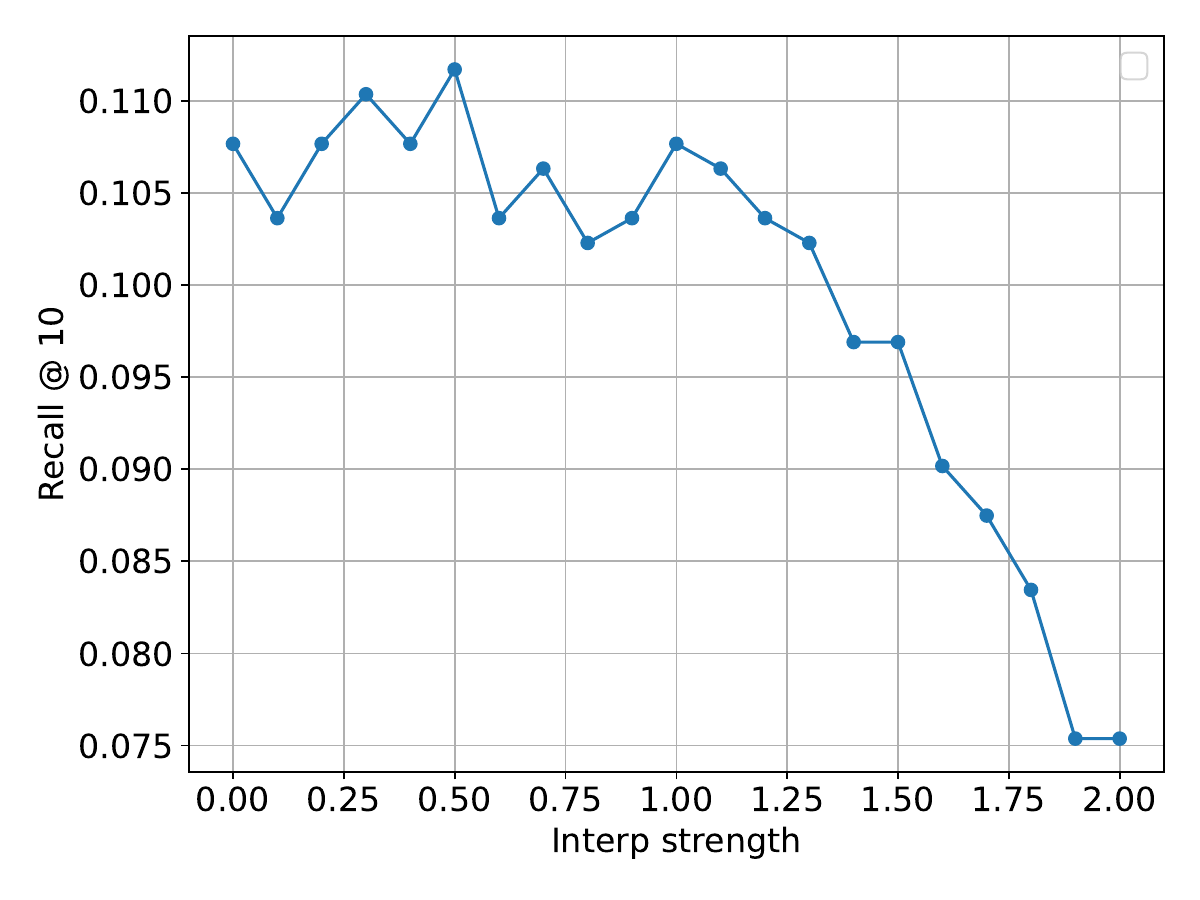}
    \caption{Recall@10 vs. text steering strengths (left) and spherical interpolation strengths (right) on the MusicCaps dataset. Image guidance strength is fixed at 19.0.}
    \label{fig:recall-vs-text-gs-musiccaps-r10}
\end{figure}

\begin{figure}[ht]
    \centering
    \includegraphics[width=0.48\linewidth]{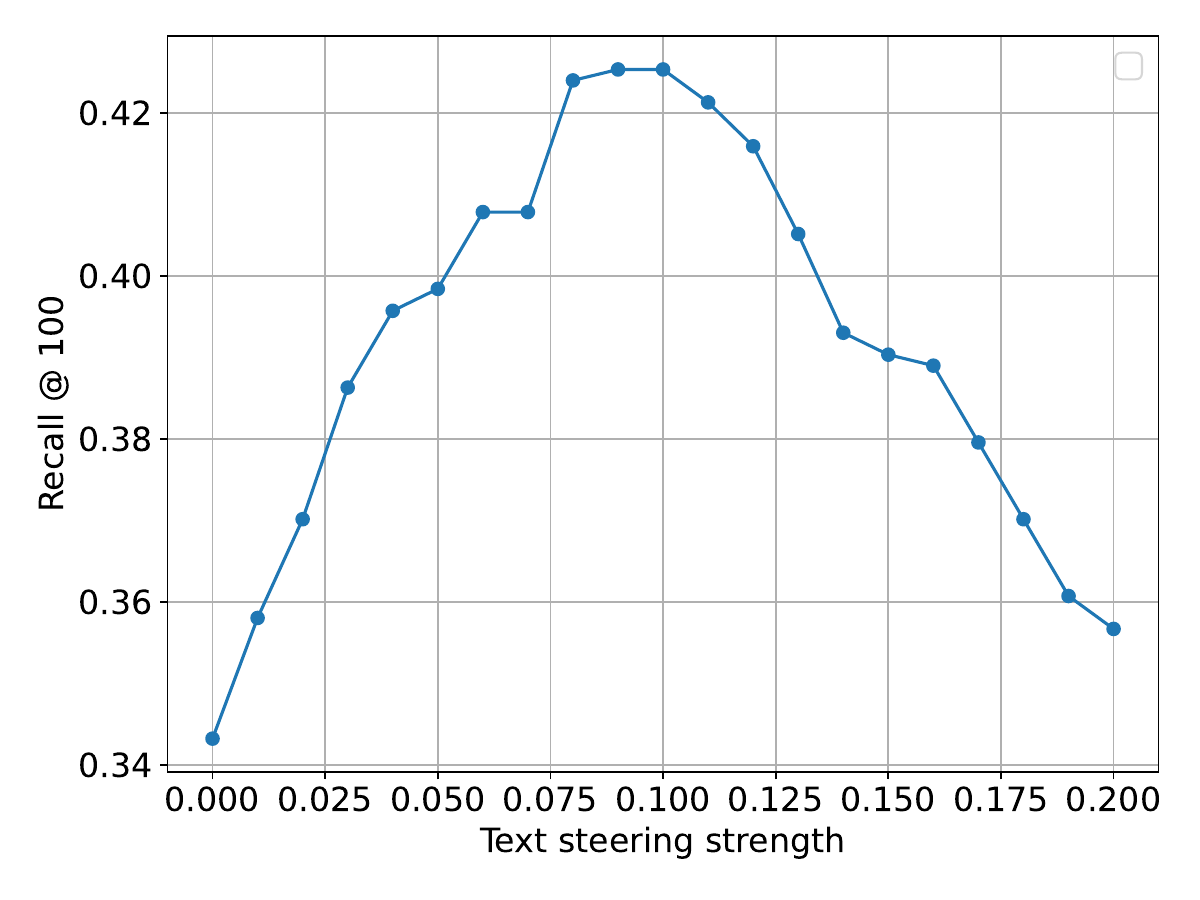}
    \includegraphics[width=0.48\linewidth]{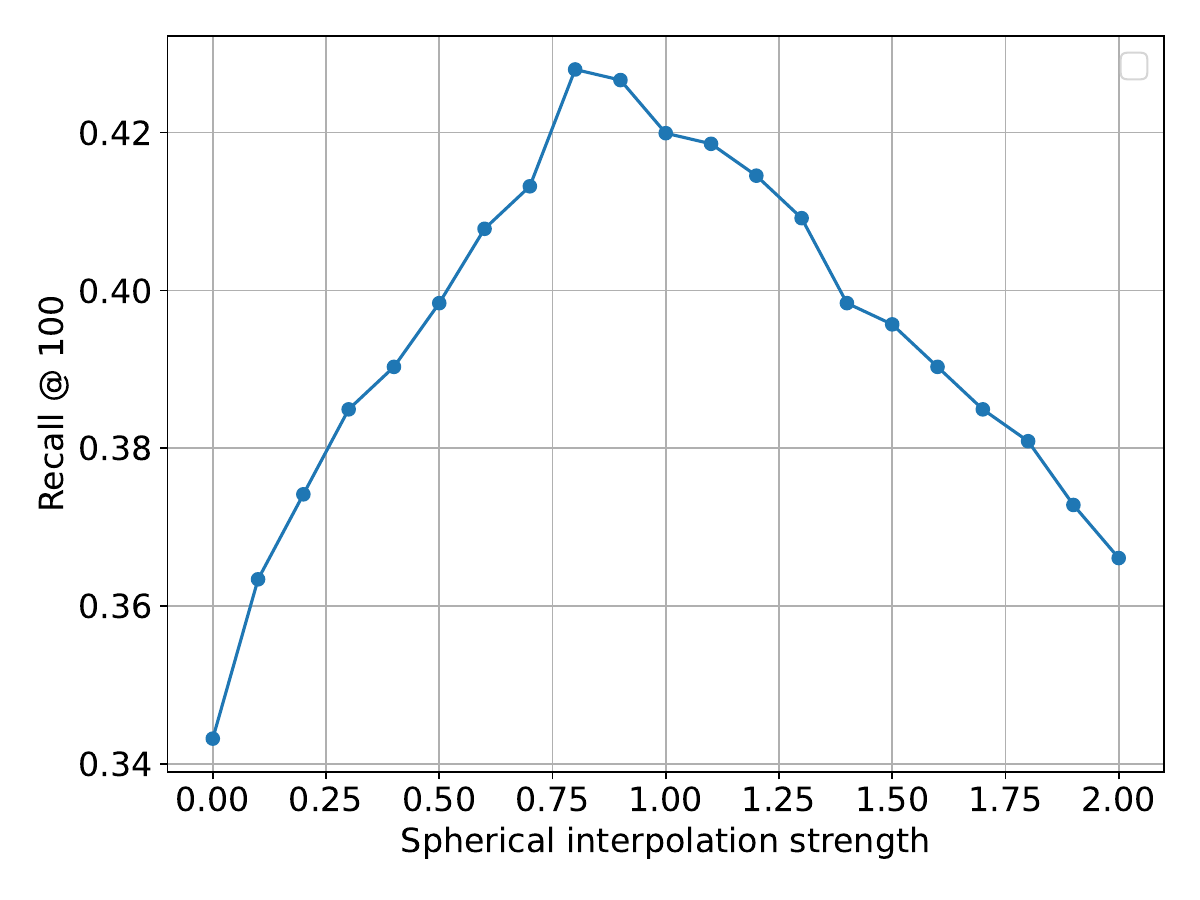}
    \caption{Recall@100 vs. text steering strengths (left) and spherical interpolation strengths (right) on the MusicCaps dataset. Image guidance strength is fixed at 19.0.}
    \label{fig:recall-vs-text-gs-musiccaps-r100}
\end{figure}

For quality of generated music embeddings, we show FMD and MISCS plots as functions of the image CFG strengths in Figure~\ref{fig:fmd-gs-musiccaps} and~\ref{fig:fmd-gs-melbench}. Figure~\ref{fig:recall-vs-gs-musiccaps} shows how the recall of music retrieval is affected by the image CFG strength. Figure~\ref{fig:recall-vs-text-gs-musiccaps-r10} and~\ref{fig:recall-vs-text-gs-musiccaps-r100} show how the recall is affected by the text steering or spherical interpolation strengths.



\end{document}